\documentclass[11pt,a4paper]{article}

\usepackage[utf8]{inputenc}
\usepackage{amsmath,latexsym}
\usepackage{xcolor}
\usepackage{graphicx}

\usepackage{axodraw2}

\usepackage{tikz}
\usepackage[nomessages]{fp}
\usepackage{jheppub}
\usepackage{glossaries}

\usepackage{accents}
  
\newcommand\munderbar[1]{%
  \underaccent{\bar}{#1}}

\def\beq{\begin{equation}}   
\def\eeq{\end{equation}}
\def\bea{\begin{eqnarray}}  
\def\eea{\end{eqnarray}} 
\def\nn{\nonumber}
\def\r{\right} 
\def\l{\left} 
\def\f21{{}_2F_{1}}
\def\eps{\epsilon}

\def\T{\mathcal{T}}
\def\g{\gamma}
\def\D{\Delta}

\makeglossaries

\def\bubbleonek{
\raisebox{-21pt}
{
\begin{axopicture}{(78,50)(-36,-24)}
\SetScale{2}\SetWidth{0.25}\SetColor{Blue}%
\Line[arrow,arrowscale=0.4](-17,0)(-10,0)
\Line[arrow,arrowscale=0.4](10,0)(17,0) 
\CArc[arrow,clockwise,arrowscale=0.4](0,0)(10,0,180)
\CArc[arrow,clockwise,arrowscale=0.4](0,0)(10,180,360)
\Vertex(-10,0){0.75}
\Vertex(10,0){0.75}
\SetColor{Black}%
\Text(-19,0){$p$}
\Text(19,0){$p$}
\Text(1.0,13.5){$k+p$ }
\Text(1.0,-14){ $k$ }
\end{axopicture}
}
}

\def\bubbleonenum#1#2{
\raisebox{-7pt}
{
\begin{axopicture}{(30,20)(-13,-10)}
\SetScale{1}\SetColor{Blue}%
\Line(-15,0)(-10,0)
\Line(10,0)(15,0) 
\CArc(0,0)(10,0,360)
\Vertex(-10,0){1.5}
\Vertex(10,0){1.5}
\SetScale{1}\SetColor{Black}%
\Text(1.5,6.5){\tiny $#1$ \tiny}
\Text(1.5,-6.5){\tiny $#2$ \tiny}
\end{axopicture}
}
}

\def\bubbleonedotnum#1#2{
\raisebox{-7pt}
{
\begin{axopicture}{(30,20)(-13,-10)}
\SetScale{1}\SetColor{Blue}%
\Line(-15,0)(-10,0)
\Line(10,0)(15,0) 
\CArc(0,0)(10,0,360)
\Vertex(-10,0){1.5}
\Vertex(10,0){1.5}
\Vertex(0,-10){1.5}
\SetScale{1}\SetColor{Black}%
\Text(1.5,6.5){\tiny $#1$ \tiny}
\Text(1.5,-6.2){\tiny $#2$ \tiny}
\end{axopicture}
}
}

\def\bubbleonedotknum#1#2{
\raisebox{-21pt}
{
\begin{axopicture}{(78,50)(-36,-24)}
\SetScale{2}\SetWidth{0.25}\SetColor{Blue}%
\Line[arrow,arrowscale=0.4](-17,0)(-10,0)
\Line[arrow,arrowscale=0.4](10,0)(17,0) 
\CArc[arrow,clockwise,arrowscale=0.4](0,0)(10,0,90)
\CArc[arrow,clockwise,arrowscale=0.4](0,0)(10,90,180)
\CArc[arrow,clockwise,arrowscale=0.4](0,0)(10,180,360)
\Vertex(-10,0){0.75}
\Vertex(10,0){0.75}
\Vertex(0,-10){0.75}
\SetColor{Black}%
\Text(-19,0){$p$}
\Text(19,0){$p$}
\Text(1.0,13.5){$k+p$ }
\Text(1.0,-14){ $k$ }
\Text(0.7,6.5){\tiny $#1$ \tiny}
\Text(0.7,-6.5){\tiny $#2$ \tiny}
\end{axopicture}
}
}

\def\bubbleonedotdotdotnum#1#2{
\raisebox{-7pt}
{
\begin{axopicture}{(30,20)(-13,-10)}
\SetScale{1}\SetColor{Blue}%
\Line(-15,0)(-10,0)
\Line(10,0)(15,0) 
\CArc(0,0)(10,0,360)
\Vertex(-10,0){1.5}
\Vertex(10,0){1.5}
\Vertex(0,10){1.5}
\Vertex(-5,-8.5){1.5}
\Vertex(5,-8.5){1.5}
\SetScale{1}\SetColor{Black}%
\Text(1.5,5.5){\tiny $#1$ \tiny}
\Text(1.5,-5.5){\tiny $#2$ \tiny}
\end{axopicture}
}
}

\def\bubbleonedotdotnum#1#2#3{
\raisebox{-7pt}
{
\begin{axopicture}{(30,20)(-13,-10)}
\SetScale{1}\SetColor{Blue}%
\Line(-15,0)(-10,0)
\Line(10,0)(15,0) 
\CArc(0,0)(10,0,360)
\Vertex(-10,0){1.5}
\Vertex(10,0){1.5}
\Vertex(-5,8.5){1.5}
\Vertex(5,8.5){1.5}
\SetScale{1}\SetColor{Black}%
\Text(1.5,5.5){\tiny $#1$ \tiny}
\Text(1.5,-5.5){\tiny $#2$ \tiny}
\Text(-12.5,-5.5){\tiny $#3$ \tiny}
\end{axopicture}
}
}

\def\bubbleonedotupdownnum#1#2#3{
\raisebox{-7pt}
{
\begin{axopicture}{(30,20)(-13,-10)}
\SetScale{1}\SetColor{Blue}%
\Line(-15,0)(-10,0)
\Line(10,0)(15,0) 
\CArc(0,0)(10,0,360)
\Vertex(-10,0){1.5}
\Vertex(10,0){1.5}
\Vertex(0,10){1.5}
\Vertex(0,-10){1.5}
\SetScale{1}\SetColor{Black}%
\Text(1.5,5.5){\tiny $#1$ \tiny}
\Text(1.5,-5.5){\tiny $#2$ \tiny}
\Text(-12.5,-5.5){\tiny $#3$ \tiny}
\end{axopicture}
}
}

\def\triangleonenum#1#2#3{
\raisebox{-7pt}
{
\begin{axopicture}{(30,20)(-8,-10)}
\SetScale{1}\SetColor{Blue}%
\Line(0,-10)(0,10)
\Line(-5,-10)(0,-10)
\Line(-5,10)(0,10)
\Vertex(0,10){1.5}
\Vertex(0,-10){1.5}
\Vertex(17.32,0){1.5}
\Line(0,-10)(17.32,0)
\Line(0,10)(17.32,0)
\Line(17.32,0)(22.32,0)
\SetScale{1}\SetColor{Black}%
\Text(-1,0){\tiny $#1$ \tiny}
\Text(11,8){\tiny $#2$ \tiny}
\Text(11,-8){\tiny $#3$ \tiny}
\end{axopicture}
}
}

\def\triangletwonum#1#2#3#4{
\raisebox{-7pt}
{
\begin{axopicture}{(30,20)(-13,-10)}
\SetScale{1}\SetColor{Blue}%
\Line(-15,0)(-10,0)
\Line(10,0)(15,0)
\Line(0,10)(0,15) 
\CArc(0,0)(10,0,360)
\CArc(10,10)(10,180,270)
\Vertex(-10,0){1.5}
\Vertex(10,0){1.5}
\Vertex(0,10){1.5}
\SetScale{1}\SetColor{Black}%
\Text(-5,4){\tiny $#1$ \tiny}
\Text(2,1.5){\tiny $#2$ \tiny}
\Text(6,5.5){\tiny $#3$ \tiny}
\Text(0,-7){\tiny $#4$ \tiny}
\end{axopicture}
}
}

\def\bubbletwonum#1#2#3#4{
\raisebox{-7pt}
{
\begin{axopicture}{(30,20)(-13,-10)}
\SetScale{1}\SetColor{Blue}%
\Line(-15,0)(-10,0)
\Line(10,0)(15,0) 
\CArc(0,0)(10,0,360)
\CArc(10,10)(10,180,270)
\Vertex(-10,0){1.5}
\Vertex(10,0){1.5}
\Vertex(0,10){1.5}
\SetScale{1}\SetColor{Black}%
\Text(-5,4){\tiny $#1$ \tiny}
\Text(2,1.5){\tiny $#2$ \tiny}
\Text(6,5.5){\tiny $#3$ \tiny}
\Text(0,-7){\tiny $#4$ \tiny}
\end{axopicture}
}
}

\def\bubbletwonuma#1#2#3#4#5{
\raisebox{-7pt}
{
\begin{axopicture}{(30,20)(-13,-10)}
\SetScale{1}\SetColor{Blue}%
\Line(-15,0)(-10,0)
\Line(10,0)(15,0) 
\Line(0,10)(0,-10) 
\CArc(0,0)(10,0,360)
\Vertex(-10,0){1.5}
\Vertex(10,0){1.5}
\Vertex(0,10){1.5}
\Vertex(0,-10){1.5}
\SetScale{2}\SetColor{Black}%
\Text(-5,5){\tiny $#1$ \tiny}
\Text(6,5){\tiny $#2$ \tiny}
\Text(6,-5){\tiny $#3$ \tiny}
\Text(-5,-5){\tiny $#4$ \tiny}
\Text(-2,0){\tiny $#5$ \tiny}
\end{axopicture}
}
}

\def\bubbletwodotnuma#1#2#3#4{
\raisebox{-7pt}
{
\begin{axopicture}{(30,20)(-13,-10)}
\SetScale{1}\SetColor{Blue}%
\Line(-15,0)(-10,0)
\Line(10,0)(15,0)
\CArc(0,0)(10,0,360)
\CArc(10,10)(10,180,270)
\Vertex(-10,0){1}
\Vertex(10,0){1}
\Vertex(7.1,7.1){1}
\Vertex(0,10){1}
\SetScale{1}\SetColor{Black}%
\Text(-5,4){\tiny $#1$ \tiny}
\Text(2,1.5){\tiny $#2$ \tiny}
\Text(9.5,8)(315){\tiny $#3$ \tiny}
\Text(0,-7){\tiny $#4$ \tiny}
\end{axopicture}
}
}

\def\bubbletwodotnumb#1#2#3#4{
\raisebox{-7pt}
{
\begin{axopicture}{(30,20)(-13,-10)}
\SetScale{1}\SetColor{Blue}%
\Line(-15,0)(-10,0)
\Line(10,0)(15,0)
\CArc(0,0)(10,0,360)
\CArc(10,10)(10,180,270)
\Vertex(-10,0){1.5}
\Vertex(10,0){1.5}
\Vertex(-7.1,7.1){1.5}
\Vertex(0,10){1.5}
\SetScale{1}\SetColor{Black}%
\Text(-5,4){\tiny $#1$ \tiny}
\Text(2,1.5){\tiny $#2$ \tiny}
\Text(9.5,8)(315){\tiny $#3$ \tiny}
\Text(0,-7){\tiny $#4$ \tiny}
\end{axopicture}
}
}

\def\bubbletwodotnum#1#2#3#4#5{
\raisebox{-21pt}
{
\begin{axopicture}{(70,50)(-30,-24)}
\SetScale{2}\SetWidth{0.25}\SetColor{Blue}%
\Line(-15,0)(-10,0)
\Line(10,0)(15,0) 
\Line(0,10)(0,-10) 
\CArc(0,0)(10,0,360)
\Vertex(-10,0){0.75}
\Vertex(10,0){0.75}
\Vertex(0,10){0.75}
\Vertex(0,0){0.75}
\Vertex(0,-10){0.75}
\Vertex(-7.07,-7.07){0.75}
\SetScale{2}\SetColor{Black}%
\Text(-5,5){\tiny $#1$ \tiny}
\Text(6,5){\tiny $#2$ \tiny}
\Text(6,-5){\tiny $#3$ \tiny}
\Text(-5,-5){\tiny $#4$ \tiny}
\Text(-2,0){\tiny $#5$ \tiny}
\end{axopicture}
}
}

\def\bubbletwodotnumc#1#2#3{
\raisebox{-7pt}
{
\begin{axopicture}{(30,20)(-13,-10)}
\SetScale{1}\SetColor{Blue}%
\Line(-15,0)(-10,0)
\Line(10,0)(15,0) 
\Line(-10,0)(10,0) 
\CArc(0,0)(10,0,360)
\Vertex(0,0){1.5}
\Vertex(-10,0){1.5}
\Vertex(10,0){1.5}
\SetScale{1}\SetColor{Black}%
\Text(1.5,6.5){\tiny $#1$ \tiny}
\Text(1.5,-6.5){\tiny $#2$ \tiny}
\end{axopicture}
}
}

\def\bubbletwodotnume#1#2#3#4#5{
\raisebox{-7pt}
{
\begin{axopicture}{(30,20)(-13,-10)}
\SetScale{1}\SetColor{Blue}%
\Line(-15,0)(-10,0)
\Line(10,0)(15,0) 
\Line(0,10)(0,-10) 
\CArc(0,0)(10,0,360)
\Vertex(-10,0){1.5}
\Vertex(10,0){1.5}
\Vertex(0,10){1.5}
\Vertex(0,0){1.5}
\Vertex(0,-10){1.5}
\SetScale{2}\SetColor{Black}%
\Text(-5,5){\tiny $#1$ \tiny}
\Text(6,5){\tiny $#2$ \tiny}
\Text(6,-5){\tiny $#3$ \tiny}
\Text(-5,-5){\tiny $#4$ \tiny}
\Text(-2,0){\tiny $#5$ \tiny}
\end{axopicture}
}
}

\def\sunrise#1#2#3{
\raisebox{-12pt}
{
\begin{axopicture}{(30,20)(-13,-15)}
\SetScale{1}\SetColor{Blue}%
\Line(-15,0)(-10,0)
\Line(10,0)(15,0)
\Line(-10,0)(10,0) 
\CArc(0,0)(10,0,360)
\Vertex(-10,0){1.5}
\Vertex(10,0){1.5}
\SetScale{1}\SetColor{Black}%
\Text(1.5,6.5){\tiny $#1$ \tiny}
\Text(1.5,-4){\tiny $#2$ \tiny}
\Text(1.5,-14){\tiny $#3$ \tiny}
\end{axopicture}
}
}

\def\sunrisedot#1#2#3{
\raisebox{-7pt}
{
\begin{axopicture}{(25,25)(-11,-10)}
\SetScale{1}\SetColor{Blue}%
\Line(-15,0)(-10,0)
\Line(10,0)(15,0)
\Line(-10,0)(10,0) 
\CArc(0,0)(10,0,360)
\Vertex(-10,0){1.5}
\Vertex(10,0){1.5}
\Vertex(0,-10){1.5}
\SetScale{1}\SetColor{Black}%
\Text(1.5,6.5){\tiny $#1$ \tiny}
\Text(1.5,-4){\tiny $#2$ \tiny}
\Text(1.5,-14){\tiny $#3$ \tiny}
\end{axopicture}
}
}

\def\sunrisedotdot#1#2#3{
\raisebox{-7pt}
{
\begin{axopicture}{(25,25)(-11,-10)}
\SetScale{1}\SetColor{Blue}%
\Line(-15,0)(-10,0)
\Line(10,0)(15,0)
\Line(-10,0)(10,0) 
\CArc(0,0)(10,0,360)
\Vertex(-10,0){1.5}
\Vertex(10,0){1.5}
\Vertex(-4,-9){1.5}
\Vertex(4,-9){1.5}
\SetScale{1}\SetColor{Black}%
\Text(1.5,6.5){\tiny $#1$ \tiny}
\Text(1.5,-4){\tiny $#2$ \tiny}
\Text(1.5,-14){\tiny $#3$ \tiny}
\end{axopicture}
}
}

\def\sunrisethreeloopdotdot#1#2#3{
\raisebox{-12pt}
{
\begin{axopicture}{(30,20)(-13,-15)}
\SetScale{1}\SetColor{Blue}%
\Line(-15,0)(-10,0)
\Line(10,0)(15,0)
\CArc(0,-17)(20,60,120)
\CArc(0,17)(20,240,300)
\CArc(0,0)(10,0,360)
\Vertex(-4,9){1.5}
\Vertex(4,9){1.5}
\Vertex(-10,0){1.5}
\Vertex(10,0){1.5}
\SetScale{1}\SetColor{Black}%
\Text(1.5,6.5){\tiny $#1$ \tiny}
\Text(1.5,-4){\tiny $#2$ \tiny}
\Text(1.5,-14){\tiny $#3$ \tiny}
\end{axopicture}
}
}

\def\sunrisethreeloopdotdotb#1#2#3{
\raisebox{-12pt}
{
\begin{axopicture}{(30,20)(-13,-15)}
\SetScale{1}\SetColor{Blue}%
\Line(-4,9)(-5,14)
\Line(4,9)(5,14)
\CArc(0,-17)(20,60,120)
\CArc(0,17)(20,240,300)
\CArc(0,0)(10,0,360)
\Vertex(-4,9){1.5}
\Vertex(4,9){1.5}
\Vertex(-10,0){1.5}
\Vertex(10,0){1.5}
\SetScale{1}\SetColor{Black}%
\Text(1.5,6.5){\tiny $#1$ \tiny}
\Text(1.5,-4){\tiny $#2$ \tiny}
\Text(1.5,-14){\tiny $#3$ \tiny}
\end{axopicture}
}
}

\def\sunrisebubble#1#2#3{
\raisebox{-12pt}
{
\begin{axopicture}{(50,25)(-12,-15)}
\SetScale{1}\SetColor{Blue}%
\Line(-15,0)(-10,0)
\Line(30,0)(35,0)
\Line(-10,0)(10,0) 
\CArc(0,0)(10,0,360)
\CArc(20,0)(10,0,360)
\Vertex(-10,0){1.5}
\Vertex(10,0){1.5}
\Vertex(30,0){1.5}
\SetScale{1}\SetColor{Black}%
\Text(1.5,6.5){\tiny $#1$ \tiny}
\Text(1.5,-4){\tiny $#2$ \tiny}
\Text(1.5,-14){\tiny $#3$ \tiny}
\end{axopicture}
}
}

\def\sunriseselfenergy#1#2#3{
\raisebox{-12pt}
{
\begin{axopicture}{(60,40)(-15,-15)}
\SetScale{1}\SetColor{Blue}%
\Line(-30,0)(-20,0)
\Line(20,0)(30,0)
\Line(-20,0)(20,0) 
\CArc(0,0)(20,0,360)

\CCirc(0,17){10}{Blue}{White}
\Line(-10,17.5)(10,17.5) 

\Vertex(-10,17){1.5}
\Vertex(10,17){1.5}
\Vertex(-20,0){1.5}
\Vertex(20,0){1.5}
\SetScale{1}\SetColor{Black}%
\Text(1.5,6.5){\tiny $#1$ \tiny}
\Text(1.5,-4){\tiny $#2$ \tiny}
\Text(1.5,-14){\tiny $#3$ \tiny}
\end{axopicture}
}
}

\def\bubblethreenum#1#2#3#4#5#6{
\raisebox{-7pt}
{
\begin{axopicture}{(30,20)(-13,-10)}
\SetScale{1}\SetColor{Blue}%
\Line(-15,0)(-10,0)
\Line(10,0)(15,0) 
\CArc(0,0)(10,0,360)
\CArc(10,10)(10,180,270)
\CArc(-10,-10)(10,0,90)
\Vertex(-10,0){1.5}
\Vertex(0,-10){1.5}
\Vertex(10,0){1.5}
\Vertex(0,10){1.5}
\SetScale{1}\SetColor{Black}%
\Text(-9,8){\tiny $#1$ \tiny}
\Text(6,5){\tiny $#2$ \tiny}
\Text(10,9){\tiny $#3$ \tiny}
\Text(10,-9){\tiny $#4$ \tiny}
\Text(-9,-9){\tiny $#5$ \tiny}
\Text(-4,-5){\tiny $#6$ \tiny}
\end{axopicture}
}}

\def\bubblethreebnum#1#2#3#4#5{
\raisebox{-21pt}
{
\begin{axopicture}{(70,50)(-30,-24)}
\SetScale{2}\SetWidth{0.25}\SetColor{Blue}%
\Line(-15,0)(-10,0)
\Line(10,0)(15,0) 
\Line(10,0)(-10,0) 
\CArc(0,0)(10,0,360)
\CArc(10,10)(10,180,270)
\Vertex(-10,0){0.75}
\Vertex(10,0){0.75}
\Vertex(0,10){0.75}
\SetColor{Black}%
\Text(-5,5){\tiny $#1$ \tiny}
\Text(2,2){\tiny $#2$ \tiny}
\Text(6,6){\tiny $#3$ \tiny}
\Text(1,-2){\tiny $#4$ \tiny}
\Text(1,-8){\tiny $#5$ \tiny}
\end{axopicture}
}}

\def\budotdotnum#1#2#3#4#5#6{
\raisebox{-7pt}
{
\begin{axopicture}{(30,20)(-13,-10)}
\SetScale{1}\SetColor{Blue}%
\Line(-15,0)(-10,0)
\Line(10,0)(15,0) 
\Line(10,0)(-10,0) 
\Line(0,10)(0,0) 
\CArc(0,0)(10,0,360)
\Vertex(-10,0){1.5}
\Vertex(10,0){1.5}
\Vertex(0,5){1.5}
\Vertex(7.07,7.07){1.5}
\Vertex(0,10){1.5}
\Vertex(0,0){1.5}
\SetScale{2}\SetColor{Black}%
\Text(-5,5){\tiny $#1$ \tiny}
\Text(2,2){\tiny $#2$ \tiny}
\Text(6,6){\tiny $#3$ \tiny}
\Text(1,-2){\tiny $#4$ \tiny}
\Text(1,-8){\tiny $#5$ \tiny}
\Text(1,-8){\tiny $#6$ \tiny}
\end{axopicture}
}}

\def\bubbleonecut#1#2#3{
\raisebox{-7pt}
{
\begin{axopicture}{(30,25)(-13,-10)}
\SetScale{1}\SetColor{Blue}%
\Line(-15,0)(-10,0)
\CArc(0,0)(10,0,360)
\Vertex(3,9){1.5}
\Vertex(-3,9){1.5}
\Vertex(-10,0){1.5}
\Vertex(10,0){1.5}
\Line(10,0)(15,0) 
\SetScale{1}\SetColor{Black}%
\Text(1.5,13.5){\tiny $#3$ \tiny}
\Text(1.5,6.5){\tiny $#1$ \tiny}
\Text(1.5,-6.5){\tiny $#2$ \tiny}
\end{axopicture}
}
}

\def\bubbletwocutbasic#1#2#3#4{
\raisebox{-7pt}
{
\begin{axopicture}{(50,25)(-13,-10)}
\SetScale{1}\SetColor{Blue}%
\Line(-15,0)(-10,0)
\CArc(0,0)(10,0,360)
\Vertex(-10,0){1.5}
\Vertex(10,0){1.5}
\CArc(20,0)(10,0,360)
\Vertex(30,0){1.5}
\Line(30,0)(35,0) 
\SetScale{1}\SetColor{Black}%
\Text(1.5,6.5){\tiny $#1$ \tiny}
\Text(1.5,-6.5){\tiny $#2$ \tiny}
\Text(21.5,6.5){\tiny $#3$ \tiny}
\Text(21.5,-6.5){\tiny $#4$ \tiny}
\end{axopicture}
}
}

\def\bubbletwocut#1#2#3#4#5#6{
\raisebox{-7pt}
{
\begin{axopicture}{(50,25)(-13,-10)}
\SetScale{1}\SetColor{Blue}%
\Line(-15,0)(-10,0)
\CArc(0,0)(10,0,360)
\Vertex(3,9){1.5}
\Vertex(-3,9){1.5}
\Vertex(-10,0){1.5}
\Vertex(10,0){1.5}
\CArc(20,0)(10,0,360)
\Vertex(23,9){1.5}
\Vertex(17,9){1.5}
\Vertex(30,0){1.5}
\Line(30,0)(35,0) 
\SetScale{1}\SetColor{Black}%
\Text(1.5,13.5){\tiny $#5$ \tiny}
\Text(1.5,6.5){\tiny $#1$ \tiny}
\Text(1.5,-6.5){\tiny $#2$ \tiny}
\Text(21.5,13.5){\tiny $#6$ \tiny}
\Text(21.5,6.5){\tiny $#3$ \tiny}
\Text(21.5,-6.5){\tiny $#4$ \tiny}
\end{axopicture}
}
}

\def\bubblethreecut#1#2#3#4#5#6#7#8#9{
\raisebox{-7pt}
{
\begin{axopicture}{(67,25)(-13,-10)}
\SetScale{1}\SetColor{Blue}%
\Line(-15,0)(-10,0)
\CArc(0,0)(10,0,360)
\Vertex(3.7,9){1.5}
\Vertex(-3.7,9){1.5}
\Vertex(-10,0){1.5}
\Vertex(10,0){1.5}
\CArc(20,0)(10,0,360)
\Vertex(30,0){1.5}
\CArc(40,0)(10,0,360)
\Vertex(43.7,9){1.5}
\Vertex(36.3,9){1.5}
\Vertex(50,0){1.5}
\Line(50,0)(55,0) 
\SetScale{1}\SetColor{Black}%
\Text(1.5,13.5){\tiny $#7$ \tiny}
\Text(1.5,6.5){\tiny $#1$ \tiny}
\Text(1.5,-6.5){\tiny $#2$ \tiny}
\Text(21.5,13.5){\tiny $#8$ \tiny}
\Text(21.5,6.5){\tiny $#3$ \tiny}
\Text(21.5,-6.5){\tiny $#4$ \tiny}
\Text(41.5,6.5){\tiny $#5$ \tiny}
\Text(41.5,-6.5){\tiny $#6$ \tiny}
\Text(41.5,13.5){\tiny $#9$ \tiny}
\end{axopicture}
}
}

\def\bubblethreecuta#1#2#3#4#5#6#7#8#9{
\raisebox{-7pt}
{
\begin{axopicture}{(67,25)(-13,-10)}
\SetScale{1}\SetColor{Blue}%
\Line(-15,0)(-10,0)
\CArc(0,0)(10,0,360)
\Vertex(3.7,9){1.5}
\Vertex(-3.7,9){1.5}
\Vertex(-10,0){1.5}
\Vertex(10,0){1.5}
\CArc(20,0)(10,0,360)
\Vertex(30,0){1.5}
\CArc(40,0)(10,0,360)
\Vertex(23.7,9){1.5}
\Vertex(16.3,9){1.5}
\Vertex(50,0){1.5}
\Line(50,0)(55,0) 
\SetScale{1}\SetColor{Black}%
\Text(1.5,13.5){\tiny $#7$ \tiny}
\Text(1.5,6.5){\tiny $#1$ \tiny}
\Text(1.5,-6.5){\tiny $#2$ \tiny}
\Text(1.5,-6.5){\tiny $#2$ \tiny}
\Text(21.5,13.5){\tiny $#8$ \tiny}
\Text(21.5,6.5){\tiny $#3$ \tiny}
\Text(21.5,-6.5){\tiny $#4$ \tiny}
\Text(41.5,6.5){\tiny $#5$ \tiny}
\Text(41.5,-6.5){\tiny $#6$ \tiny}
\Text(41.5,13.5){\tiny $#9$ \tiny}
\end{axopicture}
}
}

\def\tonebubbledotnum#1#2{
\raisebox{-7pt}
{
\begin{axopicture}{(30,20)(-13,-10)}
\SetScale{1}\SetColor{Blue}%
\Line(-15,0)(-10,0)
\Line(10,0)(15,0)
\Line(0,-10)(0,10)
\CArc(0,0)(10,0,360)
\CArc(5,10)(5,180,310)
\Vertex(8.0,6.0){1.5}
\Vertex(0,10){1.5}
\Vertex(0,0){1.5}
\Vertex(0,-10){1.5}
\Vertex(-10,0){1.5}
\Vertex(10,0){1.5}
\SetScale{1}\SetColor{Black}%
\Text(8,12){\tiny $#1$ \tiny}
\Text(-9,-10){\tiny $#2$ \tiny}
\end{axopicture}
}
}

\def\bubbleirdisconnected#1#2#3{
\raisebox{-21pt}
{
\begin{axopicture}{(70,50)(-30,-24)}
\SetScale{2}\SetWidth{0.25}\SetColor{Blue}%
\Line(-15,0)(-10,0)
\Line(10,0)(15,0) 
\CArc(0,0)(10,0,360)
\CArc(-6.8,8.1)(4.5,240,380)
\CArc(6.8,8.1)(4.5,170,300)
\Vertex(-10,0){0.75}
\Vertex(10,0){0.75}
\Vertex(-8.9,4.2){0.75}
\Vertex(8.9,4.2){0.75}
\Vertex(-2.7,9.5){0.75}
\Vertex(2.7,9.5){0.75}
\SetColor{Black}%
\Text(-10.5,3){\tiny $#1$ \tiny}
\Text(0,12){\tiny $#2$ \tiny}
\Text(12,3){\tiny $#3$ \tiny}
\end{axopicture}
}
}

\def\bubbleirconnected#1{
\raisebox{-21pt}
{
\begin{axopicture}{(70,50)(-30,-24)}
\SetScale{2}\SetWidth{0.25}\SetColor{Blue}%
\Line(-15,0)(-10,0)
\Line(10,0)(15,0) 
\CArc(0,0)(10,0,360)
\CArc(3,10)(4.5,180,310)
\CArc(10,4.5)(4.5,130,270)
\Vertex(-10,0){0.75}
\Vertex(10,0){0.75}
\Vertex(-8.9,4.2){0.75}
\Vertex(6.8,7.2){0.75}
\Vertex(-1.5,9.8){0.75}
\Vertex(-6,8){0.75}
\SetColor{Black}%
\Text(-10,7.7){\tiny $#1$ \tiny}
\end{axopicture}
}
}

\def\IRonenum#1{
\raisebox{-7pt}
{
\begin{axopicture}{(10,20)(-3,-10)}
\SetScale{1}\SetColor{RedViolet}%
\Line(0,-10)(0,10)
\CCirc(0,-10){1.5}{RedViolet}{White}
\Vertex(0,0){1.5}
\CCirc(0,10){1.5}{RedViolet}{White}
\SetScale{1}\SetColor{Black}%
\Text(6,0){\tiny $#1$ \tiny}
\end{axopicture}
}
}

\def\IRonedotnum#1{
\raisebox{-7pt}
{
\begin{axopicture}{(15,20)(-5,-10)}
\SetScale{1}\SetColor{RedViolet}%
\Line(0,-10)(0,10)
\CCirc(0,-10){1.5}{RedViolet}{White}
\Vertex(0,3.33){1.5}
\Vertex(0,-3.33){1.5}
\CCirc(0,10){1.5}{RedViolet}{White}
\SetScale{1}\SetColor{Black}%
\Text(7,0){\tiny $#1$ \tiny}
\end{axopicture}
}
}

\def\IRtwonum#1#2#3{
\raisebox{-7pt}
{
\begin{axopicture}{(30,20)(-13,-10)}
\SetScale{1}\SetColor{RedViolet}%
\Line(0,-10)(0,10)
\Vertex(0,0){1.5}
\Vertex(0,10){1.5}
\Bezier(-10,-10)(-10,17)(10,17)(10,-10)
\CCirc(-10,-10){1.5}{RedViolet}{White}
\CCirc(10,-10){1.5}{RedViolet}{White}
\CCirc(0,-10){1.5}{RedViolet}{White}
\SetScale{1}\SetColor{Black}%
\Text(-6,-3){\tiny $#1$ \tiny}
\Text(5,-3){\tiny $#2$ \tiny}
\Text(14,-3){\tiny $#3$ \tiny}
\end{axopicture}
}
}

\def\IRtwodotnum#1#2#3{
\raisebox{-7pt}
{
\begin{axopicture}{(30,20)(-13,-10)}
\SetScale{1}\SetColor{RedViolet}%
\Line(0,-10)(0,10)
\Vertex(0,0){1.5}
\Vertex(0,10){1.5}
\Vertex(8.5,0){1.5}
\Bezier(-10,-10)(-10,17)(10,17)(10,-10)
\CCirc(-10,-10){1.5}{RedViolet}{White}
\CCirc(10,-10){1.5}{RedViolet}{White}
\CCirc(0,-10){1.5}{RedViolet}{White}
\SetScale{1}\SetColor{Black}%
\Text(-6,-3){\tiny $#1$ \tiny}
\Text(5,-3){\tiny $#2$ \tiny}
\Text(14,-3){\tiny $#3$ \tiny}
\end{axopicture}
}
}

\def\IRdotbubblenum#1#2#3{
\raisebox{-7pt}
{
\begin{axopicture}{(40,20)(-13,-10)}
\SetScale{1}\SetColor{RedViolet}%
\Line(-10,0)(10,0)
\CCirc(-10,0){1.5}{RedViolet}{White}
\CCirc(15,0){5}{RedViolet}{White}
\CCirc(20,0){1.5}{RedViolet}{White}
\Vertex(0,0){1.5}
\Vertex(10,0){1.5}
\Text(1,-5){\tiny $#1$ \tiny}
\Text(16,8){\tiny $#2$ \tiny}
\Text(16,-8){\tiny $#3$ \tiny}
\end{axopicture}
}
}

\def\vaconenum#1{
\raisebox{-7pt}
{
\begin{axopicture}{(25,20)(-11,-10)}
\SetScale{1}\SetColor{Blue}%
\CArc(0,0)(10,0,360)
\Vertex(-10,0){1.5}
\Vertex(10,0){1.5}
\SetScale{1}\SetColor{Black}%
\Text(1.5,-6.5){\tiny $#1$ \tiny}
\end{axopicture}
}
}

\def\vaconenumt#1{
\raisebox{-7pt}
{
\begin{axopicture}{(25,20)(-11,-10)}
\SetScale{1}\SetColor{Blue}%
\CArc(0,0)(10,0,360)
\Vertex(-10,0){1.5}
\Vertex(10,0){1.5}
\Vertex(0,10){1.5}
\SetScale{1}\SetColor{Black}%
\Text(1.5,-14){\tiny $#1$ \tiny}
\end{axopicture}
}
}

\def\vaconenumtt#1{
\raisebox{-7pt}
{
\begin{axopicture}{(25,25)(-11,-10)}
\SetScale{1}\SetColor{Blue}%
\CArc(0,0)(10,0,360)
\Vertex(-10,0){1.5}
\Vertex(10,0){1.5}
\Vertex(0,10){1.5}
\Vertex(0,-10){1.5}
\SetScale{1}\SetColor{Black}%
\Text(1.5,-14){\tiny $#1$ \tiny}
\end{axopicture}
}
}

\def\vacsunrise#1#2#3{
\raisebox{-7pt}
{
\begin{axopicture}{(30,20)(-13,-10)}
\SetScale{1}\SetColor{Blue}%
\Line(-10,0)(10,0)
\CArc(0,0)(10,0,360)
\Vertex(-10,0){1.5}
\Vertex(10,0){1.5}
\SetScale{1}\SetColor{Black}%
\Text(1.5,6.5){\tiny $#1$ \tiny}
\Text(1.5,-4){\tiny $#2$ \tiny}
\Text(1.5,-14){\tiny $#3$ \tiny}
\end{axopicture}
}
}

\def\vacsunrisedot#1#2#3{
\raisebox{-7pt}
{
\begin{axopicture}{(25,20)(-11,-10)}
\SetScale{1}\SetColor{Blue}%
\Line(-10,0)(10,0) 
\CArc(0,0)(10,0,360)
\Vertex(-10,0){1.5}
\Vertex(10,0){1.5}
\Vertex(0,-10){1.5}
\SetScale{1}\SetColor{Black}%
\Text(1.5,6.5){\tiny $#1$ \tiny}
\Text(1.5,-4){\tiny $#2$ \tiny}
\Text(1.5,-14){\tiny $#3$ \tiny}
\end{axopicture}
}
}

\def\vacsunrisedotdot#1#2#3{
\raisebox{-7pt}
{
\begin{axopicture}{(25,20)(-11,-10)}
\SetScale{1}\SetColor{Blue}%
\Line(-10,0)(10,0) 
\CArc(0,0)(10,0,360)
\Vertex(-10,0){1.5}
\Vertex(10,0){1.5}
\Vertex(-4,-9){1.5}
\Vertex(4,-9){1.5}
\SetScale{1}\SetColor{Black}%
\Text(1.5,6.5){\tiny $#1$ \tiny}
\Text(1.5,-4){\tiny $#2$ \tiny}
\Text(1.5,-14){\tiny $#3$ \tiny}
\end{axopicture}
}
}

\def\vacsunrisedotdotb#1#2#3{
\raisebox{-12pt}
{
\begin{axopicture}{(20,25)(-10,-15)}
\SetScale{1}\SetColor{Blue}%
\Line(-10,0)(10,0) 
\CArc(0,0)(10,0,360)
\Vertex(-10,0){1.5}
\Vertex(10,0){1.5}
\Vertex(0,0){1.5}
\Vertex(0,-10){1.5}
\SetScale{1}\SetColor{Black}%
\Text(1.5,6.5){\tiny $#1$ \tiny}
\Text(1.5,-4){\tiny $#2$ \tiny}
\Text(1.5,-14){\tiny $#3$ \tiny}
\end{axopicture}
}
}

\def\vacsunrisedotdotdot#1#2#3{
\raisebox{-7pt}
{
\begin{axopicture}{(25,20)(-11,-10)}
\SetScale{1}\SetColor{Blue}%
\Line(-10,0)(10,0) 
\CArc(0,0)(10,0,360)
\Vertex(-10,0){1.5}
\Vertex(10,0){1.5}
\Vertex(0,10){1.5}
\Vertex(-4,-9){1.5}
\Vertex(4,-9){1.5}
\SetScale{1}\SetColor{Black}%
\Text(1.5,13){\tiny $#1$ \tiny}
\Text(1.5,-4){\tiny $#2$ \tiny}
\Text(1.5,-14){\tiny $#3$ \tiny}
\end{axopicture}
}
}

\def\vactwonum#1#2#3#4{
\raisebox{-7pt}
{
\begin{axopicture}{(30,20)(-13,-10)}
\SetScale{1}\SetColor{Blue}%
\CArc(0,0)(10,0,360)
\CArc(10,10)(10,180,270)
\Vertex(-10,0){1.5}
\Vertex(10,0){1.5}
\Vertex(0,10){1.5}
\SetScale{1}\SetColor{Black}%
\Text(-5,4){\tiny $#1$ \tiny}
\Text(2,1.5){\tiny $#2$ \tiny}
\Text(6,5.5){\tiny $#3$ \tiny}
\Text(0,-7){\tiny $#4$ \tiny}
\end{axopicture}
}
}

\def\vacthreenum#1#2#3#4#5#6{
\raisebox{-7pt}
{
\begin{axopicture}{(30,20)(-13,-10)}
\SetScale{1}\SetColor{Blue}%
\CArc(0,0)(10,0,360)
\CArc(10,10)(10,180,270)
\CArc(-10,-10)(10,0,90)
\Vertex(-10,0){1.5}
\Vertex(0,-10){1.5}
\Vertex(10,0){1.5}
\Vertex(0,10){1.5}
\SetScale{1}\SetColor{Black}%
\Text(-9,8){\tiny $#1$ \tiny}
\Text(6,5){\tiny $#2$ \tiny}
\Text(10,9){\tiny $#3$ \tiny}
\Text(10,-9){\tiny $#4$ \tiny}
\Text(-9,-9){\tiny $#5$ \tiny}
\Text(-4,-5){\tiny $#6$ \tiny}
\end{axopicture}
}}

\def\vactwoloopdotnum#1#2#3{
\raisebox{-7pt}
{
\begin{axopicture}{(30,20)(-13,-10)}
\SetScale{1}\SetColor{Blue}%
\CArc(0,0)(10,0,360)
\Line(-10,0)(10,0)
\Vertex(-10,0){1.5}
\Vertex(10,0){1.5}
\Vertex(0,-10){1.5}
\SetScale{1}\SetColor{Black}%
\Text(1.5,14){\tiny $#1$ \tiny}
\Text(1.5,3){\tiny $#2$ \tiny}
\Text(1.5,-5){\tiny $#3$ \tiny}
\end{axopicture}
}
}

\def\vactwoloopdotupperlowernum#1#2#3{
\raisebox{-7pt}
{
\begin{axopicture}{(18,25)(-7,-10)}
\SetScale{1}\SetColor{Blue}%
\CArc(0,0)(10,0,360)
\Line(-10,0)(10,0)
\Vertex(-10,0){1.5}
\Vertex(10,0){1.5}
\Vertex(0,10){1.5}
\Vertex(0,-10){1.5}
\SetScale{1}\SetColor{Black}%
\Text(1.5,14){\tiny $#1$ \tiny}
\Text(1.5,3){\tiny $#2$ \tiny}
\Text(1.5,-5){\tiny $#3$ \tiny}
\end{axopicture}
}
}

\def\vacbubbletwocut#1#2#3#4{
\raisebox{-7pt}
{
\begin{axopicture}{(50,25)(-13,-10)}
\SetScale{1}\SetColor{Blue}%
\CArc(0,0)(10,0,360)
\Vertex(-10,0){1.5}
\Vertex(10,0){1.5}
\CArc(20,0)(10,0,360)
\Vertex(30,0){1.5}
\SetScale{1}\SetColor{Black}%
\Text(1.5,6.5){\tiny $#1$ \tiny}
\Text(1.5,-6.5){\tiny $#2$ \tiny}
\Text(21.5,6.5){\tiny $#3$ \tiny}
\Text(21.5,-6.5){\tiny $#4$ \tiny}
\end{axopicture}
}
}

\def\vecthreesunrisea{
\raisebox{-7pt}
{
\begin{axopicture}{(25,20)(-11,-10)}
\SetScale{1}\SetColor{Blue}%
\CArc(0,0)(10,0,360)
\CArc(0,-10)(14.3,45,135)
\CArc(0,10)(14.3,225,315)
\Vertex(0,10){1.5}
\Vertex(0,4){1.5}
\Vertex(-10,0){1.5}
\Vertex(10,0){1.5}
\SetScale{1}\SetColor{Black}%
\end{axopicture}
}
}

\def\vacthreebenzdot#1{
\raisebox{-7pt}
{
\begin{axopicture}{(25,20)(-11,-10)}
\SetScale{1}\SetColor{Blue}%
\CArc(0,0)(10,0,360)
\Vertex(0,10){1.5}
\Vertex(8.66,-5){1.5}
\Vertex(-8.66,-5){1.5}
\Vertex(0,0){1.5}
\Vertex(0,-10){1.5}
\Line(0,10)(0,0)
\Line(8.66,-5)(0,0)
\Line(-8.66,-5)(0,0)
\SetScale{1}\SetColor{Black}%
\Text(0,-15){\tiny $#1$ \tiny}
\end{axopicture}
}
}

\def\vacbubbletwodotdot{
\raisebox{-7pt}
{
\begin{axopicture}{(25,20)(-6,-10)}
\SetScale{1}\SetColor{Blue}%
\CArc(0,0)(5,0,360)
\Vertex(5,0){1.5}
\CArc(10,0)(5,0,360)
\Vertex(12.5,4.5){1.5}
\Vertex(12.5,-4.5){1.5}
\end{axopicture}
}
}

\def\tadonedotnum#1#2{
\raisebox{-7pt}
{
\begin{axopicture}{(25,20)(-11,-10)}
\SetScale{1}\SetColor{Blue}%
\CArc(0,5)(5,0,360)
\Line(-10,0)(10,0)
\Vertex(0,0){1.5}
\Vertex(0,10){1.5}
\SetScale{1}\SetColor{Black}%
\Text(1,13){\tiny $#1$ \tiny}
\Text(-5,-4){\tiny $#2$ \tiny}
\end{axopicture}
}
}

\def\tadonedotdotnum#1#2{
\raisebox{-7pt}
{
\begin{axopicture}{(25,20)(-11,-10)}
\SetScale{1}\SetColor{Blue}%
\CArc(0,5)(5,0,360)
\Line(-10,0)(10,0)
\Vertex(0,0){1.5}
\Vertex(3,8.5){1.5}
\Vertex(-3,8.5){1.5}
\SetScale{1}\SetColor{Black}%
\Text(1,13){\tiny $#1$ \tiny}
\Text(-5,-4){\tiny $#2$ \tiny}
\end{axopicture}
}
}

\def\vxnum#1{
\raisebox{-2pt}
{
\begin{axopicture}{(10,10)(-5,-5)}
\SetScale{1}\SetColor{Blue}%
\Line(-5,0)(5,0)
\Vertex(0,0){1.5}
\SetColor{RedViolet}
\Line(0,0)(5,5)
\Line(0,0)(-5,5)
\SetScale{1}\SetColor{Black}%
\Text(0,6.5){\tiny $#1$ \tiny}
\end{axopicture}
}
}

\def\linenum#1{
\raisebox{-7pt}
{
\begin{axopicture}{(20,10)(-10,-10)}
\SetScale{1}\SetColor{Blue}%
\Line(-15,0)(5,0)
\Vertex(-10,0){1.5}
\Vertex(0,0){1.5}
\SetScale{1}\SetColor{Black}%
\Text(0,-6.5){\tiny $#1$ \tiny}
\end{axopicture}
}
}

\def\linenumtwo#1#2{
\raisebox{-7pt}
{
\begin{axopicture}{(30,10)(-15,-10)}
\SetScale{1}\SetColor{Blue}%
\Line(-15,0)(15,0)
\Vertex(-10,0){1.5}
\Vertex(0,0){1.5}
\Vertex(10,0){1.5}
\SetScale{1}\SetColor{Black}%
\Text(-4,3){\tiny $#1$ \tiny}
\Text(6,3){\tiny $#2$ \tiny}
\end{axopicture}
}
}

\def\arctwonumup#1#2#3{
\raisebox{-7pt}
{
\begin{axopicture}{(30,20)(-13,-10)}
\SetScale{1}\SetColor{Blue}%
\CArc(0,0)(10,0,180)
\Line(-15,0)(-10,0)
\Line(10,0)(15,0)
\SetColor{RedViolet}%
\Line(-10,0)(-10,-5)
\Line(10,0)(10,-5)
\SetColor{Blue}%
\Vertex(-10,0){1.5}
\Vertex(10,0){1.5}
\SetScale{1}\SetColor{Black}%
\Text(1.5,6.5){\tiny $#1$ \tiny}
\Text(-5,-3){\tiny $#2$ \tiny}
\Text(-12,5){\tiny $#3$ \tiny}
\end{axopicture}
}
}

\def\arctwonum#1#2#3{
\raisebox{-7pt}
{
\begin{axopicture}{(30,20)(-13,-10)}
\SetScale{1}\SetColor{Blue}%
\CArc(0,0)(10,180,360)
\Line(-15,0)(-10,0)
\Line(10,0)(15,0)
\SetColor{RedViolet}%
\Line(-10,0)(-10,5)
\Line(10,0)(10,5)
\SetColor{Blue}%
\Vertex(-10,0){1.5}
\Vertex(10,0){1.5}
\SetScale{1}\SetColor{Black}%
\Text(1.5,-6.5){\tiny $#1$ \tiny}
\Text(-14,-3){\tiny $#2$ \tiny}
\Text(-5,5){\tiny $#3$ \tiny}
\end{axopicture}
}
}

\def\arcthreenum#1#2{
\raisebox{-7pt}
{
\begin{axopicture}{(30,20)(-13,-10)}
\SetScale{1}\SetColor{Blue}%
\CArc(0,0)(10,180,360)
\Line(-15,0)(-10,0)
\Line(10,0)(15,0)
\Line(0,-10)(0,-5)
\SetColor{RedViolet}%
\Line(-10,0)(-10,5)
\Line(0,-10)(0,-5)
\Line(10,0)(10,5)
\SetColor{Blue}%
\Vertex(-10,0){1.5}
\Vertex(10,0){1.5}
\Vertex(0,-10){1.5}
\SetScale{1}\SetColor{Black}%
\Text(-10,-10){\tiny $#1$ \tiny}
\Text(16,5){\tiny $#2$ \tiny}
\end{axopicture}
}
}

\def\arcdubblethreenum#1#2#3#4{
\raisebox{-7pt}
{
\begin{axopicture}{(30,20)(-13,-10)}
\SetScale{1}\SetColor{Blue}%
\CArc(0,0)(10,180,360)
\Line(-15,0)(-10,0)
\Line(10,0)(15,0)

\Line(-10,0)(10,0)
\Vertex(-10,0){1.5}
\Vertex(10,0){1.5}
\Vertex(0,0){1.5}
\SetColor{RedViolet}%
\Line(-10,0)(-10,5)
\Line(0,0)(0,5)
\Line(10,0)(10,5)
\SetColor{Blue}%
\SetScale{1}\SetColor{Black}%
\Text(-10,-10){\tiny $#1$ \tiny}
\Text(-9,8){\tiny $#2$ \tiny}
\Text(1,8){\tiny $#3$ \tiny}
\Text(11,8){\tiny $#4$ \tiny}
\end{axopicture}
}
}

\def\bubbleonenumrem#1#2{
\raisebox{-7pt}
{
\begin{axopicture}{(30,20)(-13,-10)}
\SetScale{1}\SetColor{Blue}%
\Line(-15,0)(-10,0)
\SetColor{RedViolet}
\Line(-10,0)(-5,0) 
\Line(10,0)(5,0) 
\SetColor{Blue}
\Line(10,0)(15,0) 
\CArc(0,0)(10,0,360)
\Vertex(-10,0){1.5}
\Vertex(10,0){1.5}
\SetScale{1}\SetColor{Black}%
\Text(1.5,6.5){\tiny $#1$ \tiny}
\Text(1.5,-6.5){\tiny $#2$ \tiny}
\end{axopicture}
}
}

\def\bubbleonenumremb#1#2{
\raisebox{-7pt}
{
\begin{axopicture}{(30,20)(-13,-10)}
\SetScale{1}\SetColor{Blue}%
\Line(-15,0)(-10,0)
\SetColor{RedViolet}
\Line(0,10)(0,5) 
\Line(0,-5)(0,-10) 
\SetColor{Blue}
\Line(10,0)(15,0) 
\CArc(0,0)(10,0,360)
\Vertex(-10,0){1.5}
\Vertex(10,0){1.5}
\Vertex(0,10){1.5}
\Vertex(0,-10){1.5}
\SetScale{1}\SetColor{Black}%
\Text(1.5,6.5){\tiny $#1$ \tiny}
\Text(-9,-10){\tiny $#2$ \tiny}
\end{axopicture}
}
}

\def\bubbleonenumdis#1#2{
\raisebox{-7pt}
{
\begin{axopicture}{(30,20)(-13,-10)}
\SetScale{1}\SetColor{Blue}%
\SetColor{RedViolet}
\Line(-15,0)(-10,0)
\Line(10,0)(15,0) 
\SetColor{Blue}
\CArc(0,0)(10,0,360)
\Vertex(-10,0){1.5}
\Vertex(10,0){1.5}
\SetScale{1}\SetColor{Black}%
\Text(1.5,6.5){\tiny $#1$ \tiny}
\Text(1.5,-6.5){\tiny $#2$ \tiny}
\end{axopicture}
}
}

\def\tadpolenumrem#1#2{
\raisebox{-10pt}
{
\begin{axopicture}{(15,20)(-13,-10)}
\SetScale{1}\SetColor{Blue}%
\Line(-10,0)(0,0)
\CArc(-5,5)(5,0,360)
\Vertex(-5,0){1.5}
\Vertex(-5,10){1.5}
\SetScale{1}\SetColor{Black}%
\Text(1.5,6.5){\tiny $#1$ \tiny}
\Text(1.5,-6.5){\tiny $#2$ \tiny}
\end{axopicture}
}
}

\def\bubbletwodotnumrem#1#2{
\raisebox{-7pt}
{
\begin{axopicture}{(30,20)(-13,-10)}
\SetScale{1}\SetColor{Blue}%
\Line(-15,0)(-10,0)
\Line(10,0)(15,0)
\SetColor{RedViolet}
\Line(0,-10)(0,-5)
\Line(0,10)(0,5)
\SetColor{Blue}
\CArc(0,0)(10,0,360)
\CArc(5,10)(5,180,310)
\Vertex(8.0,6.0){1.5}
\Vertex(0,10){1.5}
\Vertex(0,-10){1.5}
\Vertex(-10,0){1.5}
\Vertex(10,0){1.5}
\SetScale{1}\SetColor{Black}%
\Text(8,12){\tiny $#1$ \tiny}
\Text(-9,-10){\tiny $#2$ \tiny}
\end{axopicture}
}
}

\def\budotdotnumrem#1#2#3#4#5#6{
\raisebox{-7pt}
{
\begin{axopicture}{(30,20)(-13,-10)}
\SetScale{1}\SetColor{Blue}%
\Line(-15,0)(-10,0)
\Line(10,0)(15,0) 
\Line(10,0)(-10,0)
\SetColor{RedViolet}
\Line(0,0)(0,4) 
\Line(0,6)(0,10) 
\SetColor{Blue}
\CArc(0,0)(10,0,360)
\Vertex(-10,0){1.5}
\Vertex(10,0){1.5}
\Vertex(7.07,7.07){1.5}
\Vertex(0,10){1.5}
\Vertex(0,0){1.5}
\SetScale{2}\SetColor{Black}%
\Text(-5,5){\tiny $#1$ \tiny}
\Text(2,2){\tiny $#2$ \tiny}
\Text(6,6){\tiny $#3$ \tiny}
\Text(1,-2){\tiny $#4$ \tiny}
\Text(1,-8){\tiny $#5$ \tiny}
\Text(1,-8){\tiny $#6$ \tiny}
\end{axopicture}
}}

\def\remsunriseir#1#2#3{
\raisebox{-12pt}
{
\begin{axopicture}{(30,20)(-13,-15)}
\SetScale{1}\SetColor{Blue}%
\Line(-15,0)(-10,0)
\Line(10,0)(15,0)
\Line(-10,0)(10,0) 
\CArc(0,0)(10,0,360)
\SetColor{RedViolet}%
\Line(-10,0)(-12,5)
\Line(10,0)(12,5)
\SetColor{Blue}
\Vertex(-10,0){1.5}
\Vertex(10,0){1.5}
\SetScale{1}\SetColor{Black}%
\Text(1.5,6.5){\tiny $#1$ \tiny}
\Text(1.5,-4){\tiny $#2$ \tiny}
\Text(1.5,-14){\tiny $#3$ \tiny}
\end{axopicture}
}
}

\def\no1example{
\raisebox{-26pt}
{
\begin{axopicture}{(150,60)(10,0)}
\SetScale{0.75}
\Line(150,10)(70,70)
\SetWidth{4} \SetColor{White}
\Line(110,70)(70,10)
\Line(150,70)(110,10)
\SetWidth{0.5} \SetColor{Black}
\Line(110,70)(70,10)
\Line(150,70)(110,10)
\Line(40,40)(20,40)
\Arc(70,40)(30,90,180)
\Arc(70,40)(30,180,270)
\Line(110,70)(70,70)
\Line(70,10)(110,10)
\Line(150,70)(110,70)
\Line(110,10)(150,10)
\Arc(150,40)(30,0,90)
\Arc(150,40)(30,270,360)
\Line(200,40)(180,40)
\Vertex(40,40){1.5}
\Vertex(70,70){1.5}
\Vertex(70,10){1.5}
\Vertex(110,70){1.5}
\Vertex(110,10){1.5}
\Vertex(150,70){1.5}
\Vertex(150,10){1.5}
\Vertex(180,40){1.5}
\Vertex(82,70){1.5}
\Vertex(94,70){1.5}
\Vertex(174,58){1.5}
\Vertex(177,27){1.5}
\Vertex(166,15){1.5}
\Text(47,15)[rt]{\small $\rho \sigma$ \small}
\Text(176,18)[lt]{\small $\kappa \lambda$ \small}
\Text(128,72)[b]{\small $\mu \nu$ \small}
\Text(93,5)[t]{\small $\mu \nu$ \small}
\Text(125,5)[t]{\small $\rho \sigma$ \small}
\Text(138,40)[l]{\small $\kappa \lambda$ \small}
\end{axopicture}
}
}

\def\QCDoneloopl{
\raisebox{-9pt}
{
\begin{axopicture}{(30,20)(-10,-10)}
\SetScale{0.75}\SetColor{Black}%
\Gluon(-20,0)(-10,0){2}{1}
\Gluon(10,0)(20,0){2}{1} 
\GluonArc(0,0)(10,0,180){2}{4}
\GluonArc(0,0)(10,180,360){2}{4}
\SetColor{Red}
\Vertex(-10,0){2}
\SetColor{Black}%
\Vertex(10,0){1.5}
\SetScale{1}\SetColor{Black}%
\end{axopicture}
}
}

\def\QCDoneloopr{
\raisebox{-9pt}
{
\begin{axopicture}{(30,20)(-10,-10)}
\SetScale{0.75}\SetColor{Black}%
\Gluon(-20,0)(-10,0){2}{1}
\Gluon(10,0)(20,0){2}{1} 
\GluonArc(0,0)(10,0,180){2}{4}
\GluonArc(0,0)(10,180,360){2}{4}
\Vertex(-10,0){1.5}
\SetColor{Red}%
\Vertex(10,0){2}
\SetColor{Black}
\SetScale{1}\SetColor{Black}%
\end{axopicture}
}
}

\def\QCDtwoloop{
\raisebox{-18pt}
{
\begin{axopicture}{(40,40)(-20,-20)}
\SetScale{0.75}\SetColor{Black}%
\Gluon(-30,0)(-20,0){2}{1}
\Gluon(20,0)(30,0){2}{1} 
\Gluon(0,-20)(0,20){2}{6} 
\GluonArc(0,0)(20,0,180){2}{9}
\GluonArc(0,0)(20,180,360){2}{9}
\Vertex(-20,0){1.5}
\Vertex(20,0){1.5}
\Vertex(0,20){1.5}
\Vertex(0,-20){1.5}
\SetScale{1}\SetColor{Black}%
\end{axopicture}
}
}

\def\QCDtriangle{
\raisebox{-13pt}
{
\begin{axopicture}{(30,30)(-18,-15)}
\SetScale{0.75}\SetColor{Black}%
\Gluon(-30,0)(-20,0){2}{1}
\Gluon(-20,0)(0,20){2}{3} 
\Gluon(-20,0)(0,-20){2}{3} 
\Gluon(0,20)(0,-20){2}{4} 
\Gluon(0,-20)(10,-20){2}{1}
\Gluon(0,20)(10,20){2}{1}
\Vertex(-20,0){1.5}
\Vertex(0,20){1.5}
\Vertex(0,-20){1.5}
\SetScale{1}\SetColor{Black}%
\end{axopicture}
}
}

\def\QCDtriangleright{
\raisebox{-13pt}
{
\begin{axopicture}{(30,30)(-8,-15)}
\SetScale{0.75}\SetColor{Black}%
\Gluon(30,0)(20,0){2}{1}
\Gluon(20,0)(0,20){2}{3} 
\Gluon(20,0)(0,-20){2}{3} 
\Gluon(0,20)(0,-20){2}{4} 
\Gluon(0,-20)(-10,-20){2}{1}
\Gluon(0,20)(-10,20){2}{1}
\Vertex(20,0){1.5}
\Vertex(0,20){1.5}
\Vertex(0,-20){1.5}
\SetScale{1}\SetColor{Black}%
\end{axopicture}
}
}

\def\IRsearch{
\raisebox{-17pt}
{
\begin{axopicture}{(40,50)(-20,-20)}
\SetScale{1}\SetColor{Blue}%
\Line(-25,0)(-20,0)
\Line(20,0)(25,0) 
\CArc(0,0)(20,120,420)
\CArc(-20,-20)(20,0,90)
\Vertex(-20,0){1.5}
\Vertex(20,0){1.5}
\Vertex(20,0){1.5}
\Vertex(0,-20){1.5}
\Vertex(0,-20){1.5}
\Vertex(-9.5,17.5){1.5}
\Vertex(9.5,17.5){1.5}
\CArc(0,20)(10,0,360)
\Line(0,10)(0,30) 
\Vertex(0,10){1.5}
\Vertex(0,30){1.5}
\end{axopicture}
}}

\def\IRsearchc{
\raisebox{-17pt}
{
\begin{axopicture}{(40,50)(-20,-20)}
\SetScale{1}\SetColor{Blue}%
\SetColor{Green}
\Line(-20,0)(20,0)
\SetColor{Blue}
\CArc(0,0)(20,120,420)
\CArc(-20,-20)(20,0,90)
\Vertex(-20,0){1.5}
\Vertex(20,0){1.5}
\Vertex(20,0){1.5}
\Vertex(0,-20){1.5}
\Vertex(0,-20){1.5}
\Vertex(0,0){1.5}
\Vertex(-9.5,17.5){1.5}
\Vertex(9.5,17.5){1.5}
\CArc(0,20)(10,0,360)
\Line(0,10)(0,30) 
\Vertex(0,10){1.5}
\Vertex(0,30){1.5}
\end{axopicture}
}}

\def\IRsearchS{
\raisebox{-17pt}
{
\begin{axopicture}{(40,50)(-20,-20)}
\SetScale{1}\SetColor{Blue}%
\SetColor{Green}
\Line(-20,0)(20,0)
\SetColor{Blue}
\CArc(0,0)(20,180,360)
\DashArc(0,0)(20,120,180){1}
\DashArc(0,0)(20,0,60){1}
\CArc(-20,-20)(20,0,90)
\Vertex(-20,0){1.5}
\Vertex(20,0){1.5}
\Vertex(20,0){1.5}
\Vertex(0,0){1.5}
\Vertex(0,-20){1.5}
\Vertex(0,-20){1.5}
\Vertex(9.5,17.5){1.5}
\CArc(0,20)(10,270,90)
\DashArc(0,20)(10,90,270){1}
\Line(0,10)(0,30) 
\Vertex(0,10){1.5}
\Vertex(0,30){1.5}
\end{axopicture}
}}

\title{The R*-operation for Feynman graphs with generic numerators}

\author[a]{Franz Herzog,}
\author[a,b]{and Ben Ruijl}

\affiliation[a]{Nikhef Theory Group,\\ Science Park 105, 1098 XG Amsterdam, The Netherlands}
\affiliation[b]{Leiden University,\\ Niels Bohrweg 1, 2333 CA Leiden, The Netherlands}

\emailAdd{fherzog@nikhef.nl}
\emailAdd{benrl@nikhef.nl}

\abstract{
The $R^*$-operation by Chetyrkin, Tkachov, and Smirnov is a generalisation of the 
BPHZ $R$-operation, which subtracts both ultraviolet and infrared divergences of euclidean
Feynman graphs with non-exceptional external momenta. It can be used
to compute the divergent parts of such Feynman graphs from products of simpler Feynman graphs
of lower loops. In this paper we extend the $R^*$-operation to Feynman graphs with arbitrary numerators, including tensors. 
We also provide a novel way of defining infrared counterterms which closely resembles the definition of its ultraviolet counterpart. 
We further express both infrared and ultraviolet counterterms in terms of scaleless vacuum graphs with a logarithmic degree of divergence. 
By exploiting symmetries, integrand and integral relations, which the counterterms of scaleless vacuum graphs satisfy, 
we can vastly reduce their number and complexity. A FORM implementation of this method was used to compute the five loop 
beta function in QCD for a general gauge group. To illustrate the procedure, we compute the poles in the dimensional 
regulator of all top-level propagator graphs at five loops in four dimensional $\phi^3$ theory.
}

\begin{document}

\preprint{Nikhef 2017-011}

\maketitle


\newpage
\section{Introduction}
The appearance of divergences has perhaps been the most serious problem of Quantum Field Theory (QFT).
In particular, the treatment of ultraviolet (UV) divergences in Feynman diagrams at higher loops took many years 
to develop. Besides UV divergences, another class of divergences are encountered in the limit of vanishing 
internal and external masses. These divergences are known as collinear and soft divergences and are often 
collectively called infrared (IR) divergences. Whereas the UV and initial state-collinear divergences can 
be renormalised into physical parameters and parton densities respectively, the soft IR divergences are 
known to cancel in the sum over all Feynman diagrams \cite{Kinoshita:1962ur,Lee:1964is} contributing 
to a particular observable.

Dimensional regularisation \cite{DimReg1,DimReg2} is the prevailing way to deal with the divergences of a QFT as it conserves both Lorentz and Gauge invariance. Furthermore it regulates UV and IR divergences at the same time and 
in a similar fashion, by expressing all divergences as poles in $\eps=(4-D)/2$, where $D$ is the dimension of 
space-time. This makes it particularly convenient for analytic calculations, which are often the preferred way 
to perform multi-loop calculations. The problem to obtain a Laurent series in $\eps$ of a general higher-loop 
Feynman integral remains a challenge and is an active field of research. Driven by the need for precision, 
enormous progress has been made in the development of general methods, based on differential equations 
\cite{Kotikov:1990kg,Gehrmann:1999as,Henn:2013pwa}, Mellin-Barnes representations 
\cite{Smirnov:1999gc,Tausk:1999vh,Anastasiou:2005cb,Czakon:2005rk}, sector decomposition 
\cite{Binoth:2000ps,Roth:1996pd,Hepp:1966eg}, analytic regularisation \cite{Panzer:2014gra} and 
finite Master integrals \cite{vonManteuffel:2014qoa} using integration-by-parts (IBP) identities 
\cite{Laporta:2001dd,Anastasiou:2004vj,Studerus:2009ye,vonManteuffel:2012np, Smirnov:2008iw}. 
While these methods are general, some either require computationally expensive IBP identities or 
lead to intractable large expressions, which make analytic evaluations very challenging. 

A method to renormalise a Feynman diagram or amplitude is the Bogolioubov, Parasiuk, 
Hepp and Zimmermann (BPHZ) renormalisation scheme \cite{Bogoliubov:1957gp,Hepp:1966eg,Zimmermann:1969jj}. 
The BPHZ method is implemented by acting onto a given Feynman integral with the recursive BPHZ $R$-operation and is based solely on the graph theoretic properties of the underlying Feynman graph. 
The $R$-operation subtracts from a Feynman graph a number of counterterms which precisely capture the complicated combinatoric structure of the superficial, sub-, and overlapping UV divergences present in Feynman diagrams at arbitrary loop order. The BPHZ renormalisation prescription is not unique, in the sense that the definition of a counterterm operation in the $R$-operation can be adjusted to change to another renormalisation scheme. As such the BPHZ renormalisation can also be defined in the minimal subtraction (MS) scheme \cite{MS} of dimensional regularisation \cite{Caswell:1981ek}. Since the counterterms generated by the $R$-operation belong to a simpler class of Feynman graphs (lower loops or factorisable) than the original Feynman graph, the $R$-operation provides a prescription to compute the Laurent series in $\eps$ of any IR convergent Feynman graph. Interestingly, it has been shown the combinatorial structure of the $R$-operation gives rise to a Hopf algebra \cite{Kreimer:1997dp,Connes:1999yr}.

Unfortunately, in QFTs which contain massless particles the $R$-operation is not sufficient to render all Feynman graphs finite, due to the presence of IR divergences. A generalisation of the $R$-operation called the $R^*$-operation, was suggested more than thirty years ago by Chetyrkin, Tkachov and Smirnov in \cite{Chetyrkin:1982nn,Chetyrkin:1984xa,Smirnov:1986me}. The $R^*$-operation is capable of subtracting both the ultraviolet and the infrared divergences of euclidean \textit{non-exceptional Feynman graphs} (Feynman graphs with non-exceptional external momenta). As a result, a powerful technique known as IR rearrangement (IRR) \cite{Vladimirov:1979zm} can be applied to reroute the external momenta of a Feynman diagram and to set masses to zero, in such a way as to maximally simplify its calculation. Critically, the IRR procedure does not alter the behaviour of the superficial UV divergence of a Feynman graph, but may lead to the creation of new IR divergences. The $R^*$-operation can be used to track and subtract these extra IR divergences. In analogy to Zimmermann's forest formula as a solution to the recursive $R$-operation, a solution to the recursive $R^*$-operation can be written as a generalised infrared forest formula \cite{Chetyrkin:2017ppe,Smirnov:1986me}. Several theorems concerning the validity and correctness of the $R^*$-operation have been proven in \cite{Chetyrkin:2017ppe}. 

The $R^*$-operation has been used in a large number of milestone multi-loop Quantum Field theoretic calculations, such as the recent computation of the five-loop beta function in QCD \cite{Baikov:2016tgj,Herzog:2017ohr} or the calculation of the five- and six--loop anomalous dimensions in $\phi^4$-theory \cite{Chetyrkin:1981jq,Gorishnii:1983gp,Kleinert:1991rg,Kompaniets:2016hct,Batkovich:2016jus}. Other applications include the calculation of the hadronic R-ratio \cite{Baikov:2012zm,Baikov:2012er}, the quark mass and field anomalous dimensions at five-loop in QCD \cite{Baikov:2014qja,Baikov:2017ujl}, as well as the inclusive Higgs decay rate into a light quarks \cite{Baikov:2005rw}. The applications of the $R^*$-operation may be classified into two different types:
\begin{enumerate}
   \item[a)] the local $R^*$-method,
   \item[b)] the global $R^*$-method.
\end{enumerate}
The local $R^*$-method is based on directly applying the $R^*$-operator to individual Feynman diagrams. The global $R^*$-method seeks to globally IR rearrange an entire amplitude (a sum of appropriately weighted Feynman graphs) instead. By applying the $R^*$-operator to this decomposition, it has been possible to work out global counterterms in a number of different calculations. In fact, all applications of the $R^*$-operation in QCD beyond the three-loop level (except for \cite{Herzog:2017ohr}, which uses the method described in this work) have been based on this global approach, whereas the local $R^*$-method has been used beyond three loops only in $\phi^4$-theory. There are at least two complications that arise in a direct application of the $R^*$-operation to QCD. First, there is a performance challenge of constructing counterterms for billions of integrals at the five-loop level in the input. Second, QCD diagrams introduce irreducible numerators, which require careful treatment. The $R^*$ methods which have been advocated and used in the calculations in $\phi^4$-theory \cite{Smirnov:1986me,Phi4,Larin:2002sc,Batkovich:2014rka} are not sufficient to deal with these extra complications. 
However, the advantage of the local approach over the global approach is that the same procedure can be applied to any process on a term by term basis. Instead, the global $R^*$-method has to be worked out independently for different correlators, which was highly challenging for the five loop QCD beta function \cite{Baikov:2016tgj} and has at present only been achieved for the $SU(3)$ gauge group.

In this work we shall develop a local $R^*$-framework which allows us to compute the pole parts of non-exceptional Feynman graphs with arbitrary numerators, including tensors. 
To achieve this goal we will identify the basic building blocks of all UV counterterms as the set of scaleless vacuum tensor Feynman graphs with logarithmic superficial degree of divergence. These \textit{logarithmic tensor vacuum graphs} (LTVGs) have more symmetry than the graphs they were derived from and allow for more dot products to be rewritten, which
vastly reduces their number and their complexity. 
We will further show that all IR counterterms can be neatly extracted from the UV counterterms of LTVGs. This framework has already been used to compute the five loop beta function in QCD for a general gauge group \cite{Herzog:2017ohr}. 
In this work we present the results for the poles of all non-factorisable (those which do not factorise into products of Feynman graphs of lower loops) Feynman graphs appearing in $\phi^3$-theory in four dimensions. 
A subset of these, so far unknown, Feynman graphs is likely to provide good candidates for master integrals. 
We anticipate that these results will provide a  useful cross-check for future evaluations by alternative methods, which may also include finite parts.

The paper is organised as follows. In section \ref{divs} we will review the notions of power counting for both UV  and IR\textendash divergences which can occur in non-exceptional Feynman graphs. In section \ref{R} we will review the $R$-operation. Next, we discuss contraction anomalies and how to consistently extend the $R$-operation to Feynman graphs with generic numerators in section \ref{section:Rgeneric}. 
In section \ref{sec:rstaroperation} we review the $R^*$-operation, introduce LTVGs, derive a  new representation of the IR counterterm and discuss how to extend the $R^*$-operation to tensor Feynman graphs. In section \ref{sec:applications} we show applications at five loops. We discuss some differences between our method and the literature in section \ref{sec:discussion}. Finally, we provide conclusions and an outlook in section \ref{sec:conclusion}. A glossary is provided
in appendix \ref{sec:glossary}.

\section{Divergences in euclidean non-exceptional Feynman graphs}
\label{divs}
We impose that all Feynman graphs to be considered in the following are euclidean and will always have 
\textit{non-exceptional external momenta}. To be precise, this means that no linear combination 
of external momenta $p_1,\ldots,p_n$ vanishes:
\beq
\sum_{i\in I} p_i \neq 0, \qquad  \text{for $I$ any subset of $\{1,\ldots,n\}$}  .
\eeq
The divergences which exist in non-exceptional Feynman graphs can be classed into two types:
\begin{enumerate}
   \item [(i)]  UV divergences, related to infinite loop momentum configurations,
   \item [(ii)] IR divergences, related to vanishing loop momentum configurations.       
\end{enumerate}
In the following we shall review the basic notions of power counting for UV  and IR divergences and thereby introduce the necessary language which 
will be needed later to define the $R$- and $R^*$-operations. 

\subsection{UV divergences in Feynman graphs}
\label{UVDivs}
An important notion in UV power counting is the \textit{superficial degree of divergence} (SDD).
To compute the superficial degree of divergence $\omega(G)$ of a Feynman graph $G$, one rescales all of its independent loop momenta $k_i$ with a parameter $\lambda$, 
i.e., $k_i\to \lambda k_i$. The leading power of $\lambda$ in the limit $\lambda\to \infty$ then 
defines the superficial degree of divergence of the Feynman graph. If $\omega \ge 0$, the integral is called
superficially UV divergent.
Let us consider a simple example: 
\beq
G=\;\bubbleonek=\int \frac{d^D k}{i\pi^{D/2}} \frac{1}{k^2(k+p)^2}\,.
\eeq
Rescaling $k\to \lambda k$ and expanding to leading order around $\lambda\to\infty$ we get
\beq
\label{UVvac}
\lim_{\lambda\to\infty} G\sim\lambda^{D-4} \int \frac{d^D k}{i\pi^{D/2}} \frac{1}{k^4}\,.
\eeq
Hence in the limit $D=4$ we find $\omega(G)=0$, which is referred to as a logarithmic divergence. 
The notion of superficial degree of divergence can also be 
used to identify subdivergences, where only some loop momenta $k_i$ diverge. Let us remind the reader at this point that in any parametrisation each loop momentum always flows around in a ``loop''. Thus, the contributions to a subdivergence
are due to the propagators in the loop and the loop momentum in the numerator. This implies that UV divergences are always associated to 
one-particle-irreducible (1PI)\footnote{A graph is 1PI if it can not be separated into two by cutting any one propagator.} subgraphs of loop one or higher.
We will call any 1PI subgraph, which has a non-negative superficial degree of divergence, a \textit{UV subgraph}. As an example consider
\beq
G=\bubbletwonum{1}{2}{3}{4},\qquad \omega(G)=0\; ,
\eeq
where we have introduced the line labels $1,2,3,4$. Let us call the subgraph consisting of lines $a,b,c,\ldots$ and those vertices connecting them as $\gamma_{abc\ldots}$, such that
\beq
\gamma_{23}=\bubbleonenum{2}{3},\qquad \gamma_{124}=\triangleonenum{1}{2}{4},\qquad \gamma_{134}=\triangleonenum{1}{3}{4}
\eeq
The superficial degrees of divergence of these subgraphs are given by
\beq
\omega(\gamma_{23})=0,\qquad \omega(\gamma_{124})=-2,\qquad \omega(\gamma_{134})=-2,
\eeq
It is easy to see that $\gamma_{23}$ is nothing but an insertion of our previous example and as such it can diverge by itself. This is an instance of a \textit{UV subdivergence}. We note that the other one-loop subgraphs are finite. 
Another example is given by
\beq
G=\bubblethreenum{1}{2}{3}{4}{5}{6}\;.
\eeq
This Feynman graph has the following UV subgraphs:
\beq
\gamma_{23}=\bubbleonenum{3}{2},\qquad \gamma_{56}=\bubbleonenum{6}{5},
\eeq
with superficial degrees of divergence
\beq
\omega(\gamma_{23})=0,\qquad \omega(\gamma_{56})=0\,.
\eeq
The subgraphs $\gamma_{23}$ and $\gamma_{56}$ are \emph{strongly disjoint}: they have no common lines or vertices.
As such, the loop momenta in both graphs can diverge independently. For dimensionless vertices, the same holds for
\emph{weakly disjoint} subgraphs, which may share vertices. We will call any set of pairwise strongly disjoint UV subgraphs \textit{UV disjoint}.
As a last example let us consider
\beq
G=\bubblethreebnum{1}{2}{3}{4}{5}, \qquad \omega(G)=2\,.
\eeq
The five UV subdivergences of this Feynman graph are given by
\beq
\gamma_{23}=\bubbleonenum{3}{2},\; \gamma_{45}=\bubbleonenum{4}{5},\; \gamma_{1234}=\bubbletwonum{1}{2}{3}{4},\;
\gamma_{1235}=\bubbletwonum{1}{2}{3}{5},\; \gamma_{2345}=\bubbletwocutbasic{2}{3}{4}{5} \, ,
\eeq
with 
\beq
\omega(\gamma_{23})=\omega(\gamma_{45})=\omega(\gamma_{1234})=\omega(\gamma_{1235})=\omega(\gamma_{2345})=0\; .
\eeq
All of these subgraphs pairwise overlap: they share at least one common vertex or line. Thus, no combination of
these subgraphs can diverge independently.

\subsection{IR divergences in non-exceptional euclidean Feynman graphs}
\label{IRdivs}
In analogy to the UV one can quantify the degree of IR divergence of a Feynman graph or subgraph $G$ by introducing the notion 
of an \textit{IR superficial degree of divergence} $\tilde\omega(G)$. Let us consider 
the simple example
\beq
G=\bubbleonedotknum{1}{2}=\int \frac{d^D k}{i\pi^{D/2}} \frac{1}{(k^2)^2(k+p)^2}\; .
\eeq
We will use a small dot to indicate a vertex. A vertex which has only two edges in 
effect creates a squared propagator. 
Such squared propagators are the simplest instance of an IR divergence in non-exceptional Feynman graphs. 
We note that a graph can only have a superficial IR divergence if no external momenta flow
through it. To study the IR properties of the line with a dot, we route the external momentum through the top line
and compute its superficial degree of divergence by rescaling, $k\to\lambda k$. This time we wish to extract the leading 
power $-\tilde\omega$ of $\lambda$ in the limit $\lambda\to0$. Performing this rescaling and taking the limit we find
\beq
\label{IRvac}
\lim_{\lambda\to0}G\sim \lambda^{D-4} \frac{1}{p^2} \int \frac{d^D k}{i\pi^{D/2}} \frac{1}{k^4}\,.
\eeq
And thus we get that $\tilde\omega(G)=0$ for $D=4$, indicating that the integral diverges logarithmically in the IR. 
One interesting difference between IR and UV divergences is that IR divergences are not usually associated to 1PI subgraphs, but they are (with minor exceptions) associated to connected subgraphs, 
which, as it turns out, is why many of the features of UV divergences explained above extend to the case of 
IR divergences. The exception is always related to self-energy insertions, of which we will give an example further below. 
This case was in fact missed in the original $R^*$-paper \cite{Chetyrkin:1982nn}  
and was later corrected in \cite{Chetyrkin:1984xa}. 

We will use the notation $\gamma'_{abc\ldots}$ to identify a certain \textit{IR subgraph} containing lines $a,b,c,\ldots$ and those vertices which 
connect only to $a,b,c,\ldots$, but no other lines. 
That is, the graph $\gamma'$ does not contain any vertices through which it is connected to the remaining graph; 
such vertices will be called external to the graph $\gamma'$. We define the graph $\bar \gamma = G \backslash \gamma'$ to be the \textit{remaining graph} where the IR subgraph is deleted from $G$. Further we define the \textit{contracted vacuum graph} $\tilde\gamma = G / \bar \gamma$ by contracting $\bar \gamma$ to a point in $G$. We remark that the
UV construction is different: the contracted vacuum graph $\tilde\gamma$ contains the IR divergence, whereas the contracted graph $G/\gamma$ for a UV subgraph $\gamma$ does not contain the UV-divergence. An example is given by
\beq
\label{eq:IRonenum}
G = \bubbleonedotnum{1}{2} \quad, \gamma_{2}'=\IRonenum{2}, \quad \bar \gamma_2 = \arctwonumup{1}{2}{}, \quad \tilde \gamma_2=\vaconenum{2} \,, 
\eeq
where we have use filled dots to denote internal vertices and hollow dots to denote external vertices. 
The momenta of the IR subgraph are considered as external to the remaining graph 
and are indicated with red amputated lines in $\bar \gamma$. It is worth remarking here that the associated contracted 
vacuum graph $\tilde\gamma$ contains the same IR divergent behaviour as the original graph, even though it is scaleless and thus vanishing 
in dimensional regularisation. Indeed, the integrand of eq.(\ref{IRvac}) is nothing but the integrand of $\tilde \gamma_2$. 

Now that we have defined the appropriate notations, we describe the conditions for an IR subgraph to be \textit{IR irreducible} (IRI) \cite{Chetyrkin:1984xa}:
\begin{enumerate}
   \item[(i)] No external momentum flows into an internal vertex of $\gamma'$,
   \item[(ii)] $\gamma'$ cannot contain massive lines, 
   \item[(iii)] the associated contracted (vacuum) graph $\tilde\gamma$
   cannot contain cut-vertices
   \footnote{A cut-vertex is a vertex which when cut separates a graph into two or more disconnected subgraphs.},
   \item[(iv)] [for insertions] each connected component in the remaining graph $\bar \gamma$ should be 1PI after shrinking massive lines and welding the vertices together that have external momenta attached to it.
\end{enumerate}
Rule (i) and (ii) follow straightforwardly from the fact that such propagators are IR regulated. Rule (iii)
prevents cases such as:
\beq
G = \vactwoloopdotupperlowernum{1}{2}{3} \quad, \tilde \gamma_{13}=\vacbubbletwocut{1}{}{3}{} \, ,
\eeq
where in fact $\gamma_{1}'$ and $\gamma_{3}'$ are two separate IR subgraphs. Finally, rule (iv) treats IRI subgraphs
that appear disjoint in the diagram:
\beq
\label{eq:irdisconnected}
G = \bubbleirdisconnected{1}{2}{3} \, ,
\eeq
where $\gamma'_{12}$, $\gamma'_{13}$, and $\gamma'_{23}$ are not IRI if lines 1,2,3 are all massless propagators 
and no additional external momentum flows through them. In that case, only $\gamma'_{123}$ is IRI, which 
can easily be seen by this equivalence:
\beq
\bubbleirdisconnected{}{}{} = \bubbleirconnected{} \, .
\eeq

Let us now consider the following case,
\beq
G=\bubbleonedotdotdotnum{1}{2} \, .
\eeq
This graph contains two distinct IR subgraphs
\beq
\gamma_{1}'=\IRonenum{1},\qquad \gamma_{2}'=\IRonedotnum{2} \,,
\eeq
with the following degree of divergences:
\beq
\tilde\omega(\gamma_1')=0,\qquad \tilde\omega(\gamma_2')=2\;.
\eeq
One may wonder whether $\gamma_1'$ and $\gamma_2'$ could diverge simultaneously. 
However, since momentum conservation at each vertex demands the incoming momentum to flow through at least one of the two propagators,
only one of the momenta can vanish `at a time'. It is useful to define, in analogy to the UV, the notion of IR disjointness. 
Loosely speaking we will call any set of non-overlapping (no common lines or internal vertices) IR subgraphs 
which can diverge simultaneously \textit{IR disjoint}. 
This can be formulated more precisely as follows. A set $S'$ of IR subgraphs of $G$ is said to be IR disjoint if the following criteria are met:
\begin{enumerate}
   \item [(i)] the IR subgraphs in $S'$ are pairwise non-overlapping,
   \item [(ii)] no other IR subgraphs can be composed from any subset of IR subgraphs in $S'$,
   \item [(iii)] the remaining graph $G \setminus S'$, defined by deleting in $G$ all lines and internal vertices of all IR subgraphs in $S'$, 
   is connected.
\end{enumerate}
For the Feynman graph above, each of $\gamma_1'$ or $\gamma_2'$ then form themselves an IR disjoint set (of one element) but their union does not. 
To see this in practice let us consider the Feynman graph
\beq
G=\bubbletwodotnum{1}{2}{3}{4}{5}\, ,
\eeq
which has the following IR subgraphs,
\beq
\gamma'_4=\IRonenum{4},\quad \gamma'_5=\IRonenum{5},\qquad \gamma'_{125}=\IRtwonum{1}{5}{2},\qquad \gamma'_{345}=\IRtwodotnum{3}{4}{5}
\eeq
with superficial degree of divergences,
\beq
\tilde\omega(\gamma_4')=\tilde\omega(\gamma_5')=\tilde\omega(\gamma'_{125})=0,\qquad \tilde\omega(\gamma'_{345})=2\;.
\eeq
Let us also consider the IR disjoint sets of IR subgraphs, which can occur in this example. From momentum flow considerations one can see 
that the only possible choice of having several disconnected IR divergences occurring simultaneously
 would be $\{\gamma_4',\gamma_5'\}$. However momentum conservation (at the vertex connecting lines $3,4$ and $5$) would in this case also force propagator $3$ to vanish, 
leading to the divergence $\gamma_{345}'$, which we have already covered. In such a case we say that union of the IR subgraphs $\gamma'_3$
and $\gamma'_4$ \textit{compose} $\gamma'_{345}$. Thus, this combination is prohibited by rule (ii). 

\section{The $R$-operation in the MS-scheme}  
\label{R}
In the following we will review the $R$-operation in the MS-scheme \cite{Caswell:1981ek}.
In the MS-scheme divergences are isolated as poles in the dimensional regulator $\eps$.
It is convenient to introduce a pole operator $K$, which acting on an arbitrary meromorphic function $F(\eps)$,
with Laurent series 
\beq
F(\eps)=\sum_{n=-\infty}^\infty c_n \eps^n\,,
\eeq
will return only its poles, i.e.,
\beq
K F(\eps)=\sum_{n=1}^{\infty} \frac{c_{-n}} {\eps^n}\, .
\eeq
The $K$-operation acting on a product of meromorphic functions fulfils what is known as a \emph{Rota-Baxter algebra}:
\beq
\label{eq:RBalgebra}
K(A B) = K(A K(B)) + K(B K(A))-K(A)K(B)\;.
\eeq
The $R$-operation will make any purely UV divergent Feynman graph $G$ finite. The following equation must therefore hold in any renormalisation scheme:
\beq
K\,R\,G=0\,.
\eeq
Writing $R=1+ \delta R$, where $\delta R$ denotes the counterterms generated by $R$, we obtain:
\beq
\label{eq:kg}
K G = -K \delta R G\,. 
\eeq

\subsection{Definition of the $R$-operation in the MS-scheme}
The $R$-operation renders a Feynman graph finite by subtracting from it all of its UV divergences. 
In section \ref{UVDivs} we introduced the concept of a UV disjoint set of subgraphs. 
This concept lies at the very heart of the $R$-operation and makes up one of the two fundamental ingredients 
with which the R-operation is equipped. These are:
\begin{enumerate}
   \item [(i)]  a notion of a set of disjoint connected subgraphs, 
   \item [(ii)] a (non-unique) counterterm operation.
\end{enumerate}
Ingredient (i) informs the R-operation about the set of distinct UV divergences which may appear in the domain of loop-integration.
The R-operation then associates a counterterm to each UV disjoint set of UV subgraphs. Mathematically the R-operation is then
expressed as a sum over the different required counterterms:
\beq
RG=\sum_{S\in W(G)} \Delta(S) * G/S \,.
\eeq
Here $\Delta(S)$ denotes\footnote{This notation unfortunately conflicts with the one of \cite{Kreimer:1997dp,Connes:1999yr}, where $\Delta$ denotes the co-product and $S$ the antipode in the Hopf algebra of graphs.} the counterterm operation acting on the ``singular'' part of $G$ specified by $S$, while $G/S$ represents 
the ``non-singular'' part of $G$, constructed by contracting in $G$ all subgraphs $\gamma$ to points. $G/S$ is also
called the remaining diagram.
$S$, sometimes referred to as a \textit{spinney}, is a set of UV disjoint UV subgraphs of $G$ and 
$W(G)$,  which is sometimes referred to as a \textit{wood}, is the set of all such sets $S$, which can be constructed from the 
UV subgraphs of $G$. $W(G)$ also includes the spinney containing the full graph, i.e., $\{G\}$. 
The $*$-operation takes the role of an insertion operator in the presence of non-logarithmic divergences, but reduces to a simple
product for logarithmic divergences.
In the case of $S=\{G\}$, we obtain the trivial identity
\beq
\Delta(\{G\})*G/\{G\}=\Delta(G)*1=\Delta(G) \, ,
\eeq
to include a counterterm for the \textit{superficial divergence} of $G$. Further $W(G)$ includes the empty set $\emptyset$, 
whose counterterm is included simply to include the full graph $G$ in the sum, i.e., 
\beq
\Delta(\emptyset)*G/\emptyset=1*G=G\; .
\eeq
Up until this point, the $R$-operation is general, but we haven't defined yet what the counterterm 
operation $\Delta$ is.
Although it has to satisfy certain criteria, the counterterm operation $\Delta$ is not unique. 
This non-uniqueness is directly related to the renormalisation scheme 
and regulator dependence which is always present in any (non-finite) Quantum Field Theory.
As such, the non-uniqueness of the counterterm operation should come as no surprise. 
Nevertheless a minimal prescription, the MS-scheme \cite{MS}, can be defined in dimensional regularisation, and it is this scheme which 
we shall employ in the following. The counterterm $\Delta(S)$ must isolate the divergence associated to the disjoint singular subgraphs $\gamma$ in $S$. 
Given their disjointness it is clear that $\Delta(S)$ must factorise,
\beq
\label{eq:Deltafactorisation}
\Delta(S) = \prod_{\gamma\in S} \Delta(\gamma)\, . 
\eeq
Let us now discuss the minimal counterterm operation $\Delta(\gamma)$. One may be tempted to 
replace $\Delta(\gamma)$ with $-K\gamma$, i.e., its divergent parts in MS. 
However, $-K\gamma$ is not a local operation if $\gamma$ contains subdivergences. These subdivergences will already have been accounted for in the sum over all $S$. 
The solution to this problem is to isolate the superficial divergence of $\gamma$ by subtracting from it all its subdivergences. 
This can be achieved by using a variant of the $R$-operation, where we just omit the counterterm of the full graph $G$ in the sum. 
We will call this operation the $\bar R$-operation:
\beq
\bar R G=\sum_{S\in \bar W(G)} \Delta(S) *G/S,\quad \text{with}\quad \bar W(G)=W(G)\setminus\{G\}\; .
\eeq
Finally we must add the $K$-operator in order to comply with the MS-scheme (where no finite pieces are kept).
This leads us to the definition of the counterterm operation:
\beq
\label{eq:CTR}
\Delta(\gamma)=-K\bar R\gamma \, .
\eeq
The counterterm operation of a graph $\Delta(\gamma)$ in the MS-scheme can be shown to be a polynomial in the 
external momenta and masses of the graph $\gamma$ of homogeneous degree $\omega(\gamma)$, see e.g. \cite{Caswell:1981ek}. 
This implies that the counterterm operation can be replaced with its Taylor expanded version:
\beq
\label{eq:UVCTR1}
\Delta(\gamma) = \sum _{n=0}^{\omega(\gamma)} \;\T^{(n)}_{\{p_i\}}\Delta(\gamma) ,
\eeq
where $\{p_i\}=\{p_1,\ldots,p_n\}$ shall denote the set of external momenta of $\gamma$ and $\T^{(n)}_{\{p_i\}}$ is a Taylor expansion operator, defined in the usual sense:
\beq
\T^{(w)}_{\{p_i\}} f(\{p_i\}) =\sum_{\alpha_1+\ldots+\alpha_n=w} \;\prod_{i=1}^n \frac{(p_i \cdot \partial_{p'_i})^{\alpha_i}}{\alpha_i!}f(\{p'_i\})\bigg\rvert_{p'_i=0},\qquad \partial_{p}^\mu=\frac{\partial}{\partial p_\mu} \, .
\eeq
We can now use the fact that derivatives with respect to external momenta commute with the $\bar R$ operation,
\beq
[\partial_{p_i}^\mu,\bar R] =0\,,
\eeq
which is true as long as no IR divergences are created, see \cite{Caswell:1981ek} for a proof, to derive 
\beq
\label{eq:UVCTR}
\Delta(\gamma) = -\sum _{n=0}^{\omega(\gamma)}  K\bar R  \,\;\T^{(n)}_{\{p_i\}}\gamma\, .
\eeq
Thus, we can further simplify the expression for the UV counterterm by replacing $\gamma$ with its Taylor expanded version.
Finally this leads us to the definition of the $*$-product in the presence of higher degree divergences, 
whose task is to insert the polynomial dependence on the external momenta $\{p_i\}$ back into the contracted graph $G/S$.

We shall see some examples of the $R$-operation for massive diagrams below.

\subsection{Examples of $R$-operations}
We shall start with our trivial example from above
\bea
R \bubbleonenum{}{} &=& 1 * \bubbleonenum{}{} +\Delta\l( \bubbleonenum{}{}\r)*1 \\
&=& \bubbleonenum{}{} +\Delta\l( \bubbleonenum{}{}\r) \nn
\eea
Note that we dropped the $*$ for the standard multiplication $\cdot$ in the second line.
A less trivial example is given by 
\bea
R \bubbletwonum{1}{2}{3}{4} &=& 1*\bubbletwonum{1}{2}{3}{4}+\Delta \l(\bubbletwonum{1}{2}{3}{4}\r)*1+\Delta \l(\bubbleonenum{2}{3}\r) *\bubbleonenum{1}{4} \\
&=& \bubbletwonum{1}{2}{3}{4}+\Delta \l(\bubbletwonum{1}{2}{3}{4}\r)+\Delta \l(\bubbleonenum{2}{3}\r) \cdot\bubbleonenum{1}{4} \;.\nn
\eea
A three-loop example is given by
\bea
R\bubblethreenum{1}{2}{3}{4}{5}{6} &=& 1*\bubblethreenum{1}{2}{3}{4}{5}{6}+\Delta \l(\bubblethreenum{1}{2}{3}{4}{5}{6}\r)*1 \nn\\
&& +\Delta \l(\bubbleonenum{5}{6}\r)* \bubbletwonum{1}{2}{3}{4} +\Delta \l(\bubbleonenum{2}{3}\r)* \bubbletwonum{4}{5}{6}{1} \\
&& +\Delta \l(\bubbleonenum{5}{6}\r) \Delta \l(\bubbleonenum{2}{3}\r) *\bubbleonenum{1}{4}\nn
\eea
Assuming that all propagators are massive, we can recursively obtain the values of $\Delta$ from those of massive tadpoles:
\bea
\Delta \l(\bubbleonenum{}{}\r) &=& -K(\bubbleonenum{}{})=\Delta \l(\vaconenum{}{}\r)= -K\l( \vaconenum{}{}\r)\\
\Delta \l(\bubbletwonum{}{}{}{}\r) &=& \Delta \l(\vactwonum{}{}{}{}\r)=-K \l( \vactwonum{}{}{}{}+\Delta \l(\bubbleonenum{}{}\r)* \vaconenum{}{} \r)\\
\Delta \l( \bubblethreenum{}{}{}{}{}{} \r) &=&\Delta \l(\vacthreenum{}{}{}{}{}{}\r) = -K\bigg( \vacthreenum{}{}{}{}{}{}
+2\Delta \l(\vaconenum{}{}\r)*\vactwonum{}{}{}{}\nn\\
&&\quad \quad +\Delta \l(\vaconenum{}{}\r)\Delta \l(\vaconenum{}{}\r)*\vaconenum{}{}\bigg)
\eea

Next, we consider a massive quadratic graph, which we Taylor expand:
\bea
\Delta \sunrise{1}{2}{3} &=& -K \bar R \vacsunrise{1}{2}{3} + 2 Q^\alpha K \bar R \left[ p_3^\alpha \vacsunrisedot{1}{2}{3} \right]\\
&-&4 Q^\alpha Q^\beta K \bar R \left[ p_3^\alpha p_3^\beta \vacsunrisedotdot{1}{2}{3} \right]
+ Q^\alpha Q^\beta K \bar R \left[ g^{\alpha \beta} \vacsunrisedot{1}{2}{3} \right]
\eea

We now see the emergence of tensor diagrams. In the next chapter we provide a consistent $R$ formalism to compute those
graphs.

\section{The $R$-operation for generic Feynman graphs in the MS-scheme}
\label{section:Rgeneric}
Let us call a \textit{generic Feynman graph} any Feynman graph which contains products of vectors or scalar products of loop momenta in the numerator. 
In order to diagrammatically denote generic Feynman graphs we introduce the following Feynman rule:
\beq
\raisebox{-12pt}
{
\begin{axopicture}{(70,30)(-30,-14)}
\Line(0,0)(40,0) 
\SetScale{2}\SetColor{Black}%
\Text(10,2){\tiny$\mu_1\mu_2 \ldots \mu_n$\tiny}
\end{axopicture}
}
\;\;=\;\;\frac{k^{\mu_1}k^{\mu_2} \cdots k^{\mu_n}}{k^2}\, .
\eeq
An example of a simple generic Feynman graph is given by
\beq
p\bubbleonenum{\mu}{\mu}p=\int \frac{d^Dk}{i\pi^{D/2}} \frac{k^\mu }{k^2} \frac{(k-p)_{\mu} }{(k-p)^2}\; .
\eeq
Before diving into the subtleties related to the application of the $R$-operation to generic Feynman graphs, let us briefly consider the possible uses of such 
a formalism and consider the potential difficulties which one may encounter. First of all, it is important to understand that the $R$-operation 
in the MS-scheme that we have presented above can readily be applied to Feynman graphs containing particles of arbitrary spin: besides scalars, 
also fermions, photons, gluons, gravitinos or gravitons, etc., connected by appropriate vertices, can straightforwardly be accommodated.
Furthermore, the $R$-operation can be used to construct the renormalisation counterterms diagram by diagram, in a way completely equivalent to that of renormalising 
the operators of the corresponding Lagrangian. It is instructive to see how this would work for a simple example in QCD:
$$
R\QCDtwoloop\,=\,\QCDtwoloop\,+\Delta\bigg(\QCDtwoloop\,\bigg)+\Delta\bigg(\QCDtriangle\bigg)* \,\QCDoneloopl +\Delta\bigg(\QCDtriangleright\bigg)* \,\QCDoneloopr\, .
$$
Here the slightly fatter red dot is used to indicate the insertion point of the triangle counterterm into the remaining graph. 
We remark that the triangle counterterm is \emph{a} 3-gluon vertex: that is, it is a local operator linear in the 
three external momenta and depends on the three external colour and Lorentz indices. 

The fact that in the MS-scheme the counterterm operation $\Delta$ is based on the $K$-operation leads to what one may call \emph{contraction anomalies}. To illustrate this, let us consider the numerators created by the evaluating the Feynman rules\footnote{Performing the Lorentz contractions among vertices and propagators and evaluating traces of gamma-matrices.}, i.e.,
\beq
\QCDtwoloop\,=\sum_{k=0}^2\eps^k N_k(\{q_i,p_i\})\bubbletwonuma{}{}{}{}{}\,.
\eeq
Here $N_k(\{k_i,p_i\})$ are polynomials in the external and internal momenta, $p_i$ and $q_i$ respectively, and we have made the dependence on $\eps$ explicit.
The contraction anomaly now emerges in the following inequality:
\beq
\label{eq:QCDcontractionanomaly}
R\l(\QCDtwoloop\,\r)\neq\sum_{k=0}^2\eps^kR\l( N_k(\{q_i,p_i\})\bubbletwonuma{}{}{}{}{}\r)\,,
\eeq
and follows from the fact that the $K$-operation does not generally commute with factors of $\eps$. The right hand side will thus lead to a different result for the renormalised Feynman graph. While both sides of \eqref{eq:QCDcontractionanomaly} would still be finite, the right hand side will no longer correspond to the finite value which one should have obtained in the \emph{original} MS-scheme. Given the argument above, it is clear that the $R$-operation for a generic Feynman graph brings with it a certain arbitrariness when it comes to the definition of its renormalisation scheme. Consequently, the operation $\bar R$ cannot be used to correctly extract the UV counterterm of a QCD diagram after the contraction of the Feynman rules has been carried out. Nevertheless,
the $R$-operation can still be used after contractions to build valid counterterms which render any Feynman graph finite. In turn, these counterterms can be used to extract the poles in $\eps$ of a given Feynman graph via eq.~\eqref{eq:kg}: $K(G) = - K \delta R G$.

\subsection{Contraction anomalies and tensor reduction}
\label{section:Rtensor}
The counterterm operation $\Delta$ and the $K$-operation do not naively commute with the contraction operation, 
as was discussed in \cite[e.g.,][]{Caswell:1981ek,Speer:1978dd}. This is an intricate feature of dimensional regularisation.
In general we have
\beq
P^{\mu_1\ldots\mu_n} \Delta \left(\gamma_{\mu_1\ldots\mu_n}\right) \neq \Delta\left(P^{\mu_1\ldots\mu_n} \gamma_{\mu_1\ldots\mu_n}\right)\; ,
\eeq
where the \textit{contraction operator} $P$ may be any even rank tensor built purely from products of metric tensors $g^{\mu\nu}$:
\beq
P^{\mu_1\ldots\mu_n} = \sum_\sigma c_\sigma g^{\mu_{\sigma(1)}\mu_{\sigma(2)}} \cdots g^{\mu_{\sigma(n-1)}\mu_{\sigma(n)}}\; .
\eeq
This means that special care has to be taken to reduce counterterms of tensor graphs to counterterms of scalar diagrams. 
Let us consider some examples to clarify this issue further. 
A simple one-loop example is given by
\beq
\label{eq:one-loop tensor}
\Delta\left(\bubbleonedotnum{}{\mu\nu}\right)=-K\bubbleonedotnum{}{\mu\nu}=A g^{\mu\nu}
\eeq
where $A$ is some constant to be determined. In general one may expect that there could be a second tensor structure of kind $Q^\mu Q^\nu$, 
but this is excluded since a counterterm of logarithmic SDD cannot depend on its external momenta. 
One may be tempted to extract the value of $A$ by contracting both sides of \eqref{eq:one-loop tensor} with a projector 
$g_{\mu\nu}/D$. However, this will lead to the wrong result, precisely because contraction does not commute with the $K$-operator. 
The correct result for the coefficient $A$ is obtained by performing the tensor reduction inside the K-operation:
\beq
K\bubbleonedotnum{}{\mu\nu}=K\left( \frac{g_{\mu\nu}}{D}\bubbleonedotnum{}{\alpha\alpha} \right)
\quad \Rightarrow \quad A=- K\left( \frac{1}{D}\bubbleonenum{}{} \right) 
\eeq
Here we used that the metric tensor can be commuted with a $K$-operation as long as this does not lead to any contractions inside that $K$.
Applying a tensor reduction for a $\Delta$-operation at higher loops is more involved: a safe procedure is to first apply the counterterm operation 
recursively to obtain an expression in terms of nested $K$-operations, and then apply the tensor reduction iteratively, starting inside the most inner $K$ and then commuting 
tensors outwards. Let us consider how this works for the following simple two loop example:
\bea
\Delta\left(\bubbletwodotnuma{}{}{\mu\nu}{}\right) &=& -K\left( \bubbletwodotnuma{}{}{\mu\nu}{} \right)
+K\left( K\left(\bubbleonedotnum{}{\mu\nu}\right)\bubbleonenum{}{}\right)\nn\\
&=& g_{\mu\nu}\l[\,-K\left(\frac{1}{D} \bubbletwonum{}{}{}{} \right)+\, K\left( 
K\left(\frac{1}{D}\bubbleonenum{}{}\right)\bubbleonenum{}{}\right)\r] \, .
\eea
We see that one cannot naively extract the value of $\Delta$ of a tensor Feynman graph from that of its tensor reduced version:
\bea
\Delta\left(\bubbletwodotnuma{}{}{\mu\nu}{}\right) &\neq& g_{\mu \nu} \Delta\left(
	\frac{1}{D} \bubbletwonum{}{}{}{}{}\right) \nn\\
	&=& g_{\mu\nu}\l[\,-K\left(\frac{1}{D} \bubbletwonum{}{}{}{} \right)+\, K\left( \frac{1}{D} 
K\left(\bubbleonenum{}{}\right)\bubbleonenum{}{}\right)\r]\, .
\eea

\subsection{Contraction anomalies and counterterm factorisation}
Another kind of contraction anomaly is encountered when one considers the counterterm operation acting on a product of 
graphs. For factorising scalar graphs the following factorisation formula holds (see appendix \ref{Appendix:Cutvertexrule} for an inductive proof):
\beq
\label{eq:cutfactorisation}
\Delta(G_1G_2)=\Delta(G_1)\Delta(G_2)\; .
\eeq
The formula no longer holds when $G_1$ and $G_2$ are replaced with tensor graphs. To clarify this issue, let us consider the following example:
\beq
\label{eq:cutexamplea}
G_a=\bubblethreecut{1}{2}{3}{4}{5}{6}{\mu_1..\mu_4}{}{\mu_1..\mu_4}\quad \;\;.
\eeq
It is perhaps worth reiterating that this generic graph is rather unphysical. In \emph{physical} graphs a Lorentz index of a particular propagator 
would always be contracted with its neighbouring vertices. Any uncontracted indices that enter a counterterm in a physical Feynman graph would 
thus always be ``external'' and should be treated as commuting with respect to the $K$-operation. 
This is rather different for the graph under consideration. Its indices are in some sense internal and yet they are uncontracted.
A problem when acting the $R$-operation onto this graph becomes apparent when representing the graph in a different order:
\beq
\label{eq:cutexampleb}
G_b=\bubblethreecuta{1}{2}{5}{6}{3}{4}{\mu_1..\mu_4\;}{\;\mu_1..\mu_4}{}\;\quad\;\;.
\eeq
The difference of acting the $R$-operation on $G_a$ and $G_b$ then becomes
\bea
RG_a-RG_b &=&\l( \Delta\l(\bubbleonecut{1}{2}{\mu_1..\mu_4\;}\r)\Delta\l(\bubbleonecut{5}{6}{\mu_1..\mu_4\;}\r)-\Delta\l(\bubbletwocut{1}{2}{5}{6}{\mu_1..\mu_4\;}{\;\mu_1..\mu_4}{}\r)\r) \bubbleonenum{3}{4}\;\nn\\
&&+\Delta\l(\bubblethreecut{1}{2}{3}{4}{5}{6}{\mu_1..\mu_4}{}{\mu_1..\mu_4}\r)-\Delta\l(\bubblethreecuta{1}{2}{5}{6}{3}{4}{\mu_1..\mu_4\;}{\;\mu_1..\mu_4}{}\r)\\
&=& -(1-K)\l(\l[ (1-K)K\l(\bubbleonecut{1}{2}{\mu_1..\mu_4\;}\r)K\l(\bubbleonecut{5}{6}{\mu_1..\mu_4\;}\r)\r]\bubbleonenum{3}{4}\r)\;.\nn \\
&=& \text{finite but non-zero}\nn ,
\eea
where we used
\beq
R=(1-K)\bar R,
\eeq
which in some sense guarantees finiteness. This shows, however, that reordering of the subgraphs effectively
results in a transition to a different renormalisation scheme, 
as we anticipated in our discussion in the beginning of this chapter. The difference in schemes may be traced back to the breaking of 
\eqref{eq:cutfactorisation}, which is caused by a violation of the identity
\beq
K(K(A)K(B))=K(A)K(B)
\eeq
if $A$ and $B$ are contracted tensors. Another way to phrase this observation is that the Rota-Baxter algebra of eq.(\ref{eq:RBalgebra}) 
is not valid for contracted tensor graphs.

Even though both $RG_a$ and $RG_b$ are finite, this calculation clearly shows that the counterterm operation $\Delta$ is not invariant under 
reorderings. A possible choice to enforce the re-ordering property is to use the replacement
\beq
\label{eq:defcutfactorisation}
\Delta(G_1G_2)\to\Delta(G_1)\Delta(G_2)\; 
\eeq
A convenient way to incorporate \eqref{eq:defcutfactorisation} into the $R$-operation is to introduce a modified notion of UV disjointness. 
If one defines a spinney to consist of a set of weakly disjoint subgraphs, i.e., subgraphs which contain no common lines but may have common vertices, the right hand side of 
\eqref{eq:defcutfactorisation} is automatically produced. Apart from leading to perfectly finite results, this \emph{weakly disjoint renormalisation scheme} has the advantages that it restores invariance under re-ordering of cut-vertex connected subgraphs. It further simplifies calculations since the right hand side of \eqref{eq:defcutfactorisation} generally leads to fewer terms than the left hand side. A proof for the finiteness of the factorised renormalisation scheme is sketched for factorised graphs in appendix \ref{Appendix:CutvertexruleTensor}.

\section{The $R^*$-operation}
\label{sec:rstaroperation}
The $R^*$-operation \cite{Chetyrkin:1982nn,Chetyrkin:1984xa,Smirnov:1986me} extends the $R$-operation to the subtraction of infrared divergences of euclidean Feynman graphs with non-exceptional external momenta. The main power of the $R^*$-formalism derives from a trick known as infrared rearrangement (IRR). This trick uses the feature that the counterterm operation associated to a logarithmic superficially UV divergent Feynman graph is independent of its external momenta and masses. Infrared rearrangement allows one to compute the counterterm operation $\Delta$ 
of a graph $G$ from a simpler rearranged version of it, called $G'$. We have essentially performed IR rearrangements in the context of the standard $R$-operation, which in the presence of internal masses can be exploited to set all external momenta to zero. 
In the absence of masses in propagators, IR rearrangement may lead to infrared divergences. This is illustrated in the following 
example:
\beq
\bubbletwonum{}{}{}{} \rightarrow \sunrisedot{}{}{} \; .
\eeq
In order to subtract the newly created IR divergences, the $R$-operation must be equipped with an infrared counterterm operation. We shall review the defintion of the $R^*$-operation in the following and extend it to generic Feynman graphs. We will furthermore introduce the concept of vacuum Feynman graphs of logarithmic SDD which will take centre stage in our formalism. We will also show that this concept 
offers an alternative description of the IR counterterm operation.

\subsection{Definition of the $R^*$-operation}
In section \ref{IRdivs} we have already introduced the notions of IR subgraphs and IR disjoint sets of IR divergences. These are all the necessary notions to readily 
generalise the $R$-operation. First, we note that a given subgraph can never at the same time become IR  and UV divergent. This trivially 
implies that overlapping UV IR counterterms do not need to be considered. The $R^*$-operation acting on a given graph $G$ then takes the following form:
\beq
R^* G = \sum_{\substack{S\in W(G),S'\in W'(G)\\S \cap S' = \emptyset}}\;\widetilde\Delta (S')* \Delta(S) * G/S \setminus S'\,.
\eeq
We will explain the details of this equation in the following. The sum goes over all non-intersecting sets $S$ and $S'$, where $S$ is a \textit{UV spinney}, 
as before a set of UV disjoint UV subgraphs, and $S'$, is an \textit{IR spinney}, a set of IR disjoint IR subgraphs. As before $W(G)$ is the set of all UV spinneys $S$ of $G$, 
whereas $W'(G)$ is the set of all IR spinneys $S'$ of $G$. An efficient algorithm to construct the IR spinney is given
in appendix \ref{sec:irsubgraphsearch}.

The IR counterterm operation $\widetilde\Delta $ factorises, similarly to the UV counterterm operation $\Delta$, 
over disjoint IR subgraphs: 
\beq
\widetilde\Delta (S')=\prod_{\gamma'\in S'} \widetilde\Delta (\gamma')\,,\qquad \Delta(S)= \prod_{\gamma\in S} \Delta(\gamma) \,.
\eeq
The reduced graph $G/S \setminus S'$ is constructed by first shrinking all UV subgraphs in $S$ into points in $G$, identically as in the case of the 
$R$-operation, and then deleting in $G/S$ all the lines and vertices of all IR subgraphs in $S'$.
The UV counterterm operation $\Delta$ is defined, identically to the case of the $\bar R$-operation, to isolate the superficial UV divergence. This is achieved by subtracting from the given Feynman graph or subgraph all UV subdivergences as well as all IR divergences. The UV counterterm operation is then defined as
\beq
\Delta(G)=-K\bar R^* G\,,
\eeq
with the $\bar R^*$-operation being defined recursively via
\beq
\bar R^* G = \sum_{\substack{S\in \bar W(G),S'\in W'(G)\\S \cap S' = \emptyset}}\;\widetilde\Delta (S')* \Delta(S) * G/S \setminus S'\,.
\eeq
The proper UV wood $\bar W(G)$ (as in the case of the $\bar R$-operation) includes all UV spinneys apart from the one which consists of the graph $G$ itself.
Equivalently, we can extract the same UV counterterm from a set of \emph{massless} vacuum integrals by Taylor expanding the graph $G$ up to order $\omega(G)$ 
around all of its external momenta $\{p_i\}$ and masses $\{m_i\}$. We will call these scaleless vacuum tensor graphs of logarithmic SDD \textit{logarithmic tensor vacuum graphs} (LTVGs).
This leads us to the alternative definition in terms of LTVGs:
\beq
\Delta(G)=-K\bar R^* \mathcal{T}^{(\omega(G))}_{\{p_i,m_i\}} G\,.
\eeq
In contrast to eq.(\ref{eq:UVCTR}) we have dropped in the above all terms in the Taylor series but the massless logarithmic ones. This relies entirely on the property that in dimensional regularisation all scaleless integrals are zero. Since non-logarithmic integrals require a scale, which is absent, they must be zero. For LTVGs this statement must be understood as a cancellation of poles between IR  and UV divergences. For a massless vacuum graph $G$ the statement is thus that unless the vacuum graph is an LTVG, its UV counterterm must be set to zero. This has some far-reaching consequences: the UV counterterm of \emph{any} generic Feynman graph can \emph{always} be expressed as a counterterm of an LTVG. In this way the counterterm operation of many seemingly different graphs can easily be related to one another. In order to compute the counterterm of an LTVG, one can then always choose to compute it from a convenient single scale graph.
For example, the UV counterterm of the following two-loop vacuum, propagator or vertex Feynman graphs are all identical, whether they include massive lines or not: 
\beq
\Delta\l(\vactwonum{}{}{}{}\r)=\Delta\l(\bubbletwonum{}{}{}{} \r)=\Delta\l(\bubbletwodotnumc{}{}{}\r)=\Delta\l(\triangletwonum{}{}{}{} \r)\,.
\eeq
Such relations, when combined with rewriting dot products and a powerful graph canonicalisation algorithm that maps each isomorphic graph to the
same graph, can be used to reduce the number of unique counterterms to a comparably small number, even at five loops.

The IR counterterm operation $\widetilde\Delta $ can be defined analogously by isolating the superficial IR divergence of the IR subgraph by subtracting from it all UV divergences as well as all IR subdivergences. For this concept to make sense it is useful to recall the definition of the contracted vacuum IR subgraph which we introduced in section \ref{IRdivs}. Given some IR subgraph $G'$, and its associated contracted vacuum graph $\tilde G$, we define 
\beq
\widetilde\Delta(G')=-K\munderbar R^* \tilde G\,,
\eeq
with the $\munderbar R^*$-operation defined recursively via
\beq
\munderbar R^* \tilde G = \sum_{\substack{S\in W(\tilde G),S'\in \munderbar W'(\tilde G)\\S \cap S' = \emptyset}}\;\widetilde\Delta (S')* \Delta(S) * \tilde G/S \setminus S'\,.
\eeq
The proper IR wood $\munderbar W(G)$ then includes all IR spinneys apart from the one which consists of the (vacuum) graph $G$ itself.
We remark that for euclidean non-exceptional Feynmangraphs only vacuum graphs can actually carry superficial IR divergences. For IR divergences with higher than logarithmic SDD, the internal momenta of the IR subgraphs $\gamma'$ should be viewed as external to $G/S\setminus S'$ and $\widetilde\Delta (\gamma')$ is promoted to 
become a Taylor expansion operator acting on the remaining graph $G/S\setminus S'$. To understand the origin of this prescription, it is useful to review how the factorisation of the IR counterterm really comes to be.
This will be discussed in more detail in the next section. 

To familiarise the reader with the underlying simplicity of the procedure, which is easily lost in the formalism, 
let us for now give a few examples of the $R^*$-operation. The simplest example is given by a one-loop bubble with a dotted line:
\bea
R^* \bubbleonedotnum{}{} &=&1* 1 * \bubbleonedotnum{}{} +\widetilde\Delta  \left(\IRonenum{} \right)*1*\arctwonumup{}{}{} \\
	&=& \bubbleonedotnum{}{} +\widetilde\Delta  \left( \IRonenum{}{} \right) \linenum{}\,.\nn
\eea
Here we use red truncated lines to indicate that the \textit{infrared propagators} enter as external lines into the vertices of the remaining graph. In the second line we have evaluated the $*$-product in the IR counterterm, which for logarithmic SDD simply results in setting the external IR legs to zero. We further used the trivial identity $1*1*G=1*G=G$. We now present a slightly more complicated example at two loops:
\bea
R^*\bubbletwodotnumc{}{}{}{}=&&1*\bubbletwodotnumc{}{}{}{}+\widetilde\Delta  \left( \IRonenum{}{} \right)*\bubbleonenumrem{}{}+\widetilde\Delta  \left( \IRonenum{}{} \right)*\Delta\l(\bubbleonenum{}{}\r)*\vxnum{}\nn\\
&& +\Delta\l(\bubbleonenum{}{}\r)*\tadpolenumrem{}{}+\Delta\l(\bubbletwodotnumc{}{}{}{}\r)*1\\
=&& \bubbletwodotnumc{}{}{}{}+\widetilde\Delta  \left( \IRonenum{}{} \right) \bubbleonenum{}{}+\widetilde\Delta  \left( \IRonenum{}{} \right) \Delta\l(\bubbleonenum{}{}\r)+\Delta\l(\bubbletwodotnumc{}{}{}{}\r)\nn \, .
\eea
This example illustrates the interplay of the subtraction of IR and UV divergences. Notice that one of the counterterms for the UV subdivergence vanishes as it contains 
a massless tadpole. Below is an example containing a two-loop IR subgraph:
\bea
R^* \bubbletwodotnumb{}{}{}{}{} &=& 1 * \bubbletwodotnumb{}{}{}{}{} 
	+ \Delta \left( \bubbleonenum{}{} \right) * \bubbleonedotnum{}{} \nn\\
	&+&\widetilde\Delta  \left(\IRonenum{} \right) * \Delta \left( \bubbleonenum{}{} \right) *\arctwonum{}{}{}
	+ \widetilde\Delta  \left( \IRdotbubblenum{}{}{} \right) *\arctwonum{}{}{} \nn\\
	&=& \bubbletwodotnumb{}{}{}{}{} 
	+ \Delta \left( \bubbleonenum{}{} \right) \bubbleonedotnum{}{} \\
	&+& \widetilde\Delta  \left(\IRonenum{} \right) \Delta \left( \bubbleonenum{}{} \right) \linenum{}
	+ \widetilde\Delta  \left( \IRdotbubblenum{}{}{} \right) \linenum{} \nn\, .
\eea

\subsection{The infrared counterterm operation}
Our presentation of the IR counterterm will differ in parts from the literature \cite{Chetyrkin:2017ppe,Chetyrkin:1982nn,Chetyrkin:1984xa,Smirnov:1986me,Phi4,Larin:2002sc}, 
it is nevertheless consistent with all accounts given there, at least when it comes to the evaluation of logarithmic IR divergences in scalar theories. Let us now study the singular behaviour of the integrand 
of the graph $G$ in the limit where the momenta $k_1,\ldots,k_n$ which are contained in a given IR divergent subgraph $\gamma'$ approach zero. In order to make the singular behaviour of $G$ explicit, let us write
\beq
G(\{k_i\})=\tilde \gamma(\{k_i\}) \cdot (G\setminus\gamma') (\{k_i\}) \, ,
\eeq
where we remind the reader that $\tilde \gamma$ is the contracted IR vacuum graph (see eq.(\ref{eq:IRonenum})). Even though the IR divergence of degree $\tilde \omega$ is entirely captured by the factor $\tilde \gamma$, the remaining graph
$G\setminus\gamma'$ still depends on the momenta $\{k_i\}$ as external momenta, 
which flow into $G\setminus\gamma'$ through those 
vertices which connect it to $\gamma'$. Having made this dependence explicit, we now Taylor expand the remaining graph around 
$k_i=0$ up to and including order $k_i^{\tilde \omega}$. The singular behaviour of $G$ in the limit $k_i\to 0$ is then entirely captured by
\beq
G(\{k_i\})=\sum_{r=0}^{\tilde \omega} \tilde \gamma(\{k_i\}) \cdot \mathcal{T}^{(r)}_{\{k_i\}} (G\setminus\gamma') (\{k_i\})+\mathcal{O}(k_i^0)\; .
\eeq
To build a counterterm for $\gamma'$ we have to isolate its superficial IR divergence. We can accomplish this by introducing
the operator $K\munderbar R^*$ which will now act not only on $\tilde \gamma_i$, but also on the polynomial $k_i$-dependent 
terms which are created by the Taylor expansion. Explicitly, this means that we can identify:
\bea
\label{eq:deltaprimeaction}
\widetilde\Delta (\gamma')* G\setminus\gamma' = -\sum_{\alpha_1+..+\alpha_n=\tilde \omega} &&\;
K\munderbar R^* \bigg(\tilde \gamma(\{k_i\}) \prod_{i=1}^n k_i^{\mu_{i1}}\cdots k_i^{\mu_{i\alpha_i}} \bigg)\; \\
&&\cdot\quad \l[\bigg(\prod_{i=1}^n \frac{1}{\alpha_i!} \partial_{k'_i}^{\mu_{i1}}\cdots \partial_{k'_i}^{\mu_{i\alpha_i}}\bigg) (G\setminus\gamma') (\{k_i'\})\r]_{k'_i=0}\nn \, ,
\eea
where we have dropped all orders in the Taylor expansion other than $\tilde \omega$, since these would give rise to scaleless vacuum graphs of non-logarithmic SDD 
which vanish under the operation $K\munderbar R^*$. Furthermore, we have introduced a set of dummy Lorentz indices $\mu_{ij}$ with 
$i\in\{1,..,n\}$ and $j\in\{1,..,\alpha_i\}$ . In conclusion we can write:
\beq
\label{eq:deltaprime}
\widetilde\Delta (\gamma')=-K\munderbar R^*  \tilde\gamma\, \mathcal{T}^{(\tilde \omega(\gamma'))}_{\{k_i\}}\,.
\eeq
We see that $\widetilde\Delta (\gamma')$ is promoted to a Taylor expansion operator which acts onto the remaining graph.
This should be contrasted to the UV counterterm where the Taylor expansion operator acts on the UV subgraph and 
simply inserts the polynomial dependence of the counterterm into the remaining graph as a vertex. Thus whereas the UV counterterm is local in 
configuration space, the IR counterterm is local in momentum space. We remark that the IR counterterm is usually presented after 
integration as a sum over derivatives of the Dirac-delta function times $Z$-factors. Our representation is completely analogous to this representation, as one can easily show. However it further illuminates that the same Taylor expansion technique, which can be used to derive the UV counterterm in momentum space a la Zimmermann \cite{Zimmermann:1969jj}, can be used in complete analogy for the case of IR divergences.

From eq.(\ref{eq:deltaprimeaction}) we see that the Taylor expansion makes the IR subgraph logarithmic by multiplying $\tilde \gamma$ with monomials of IR momenta. In effect this procedure thus relates the IR counterterm operation of IR subgraphs of arbitrary SDD to IR counterterm operations of LTVGs.
Let us denote such an LTVG by $\tilde \gamma_{\log}$. Although $\widetilde\Delta  \tilde \gamma_{\log}$ is in principle a well defined operation, one never actually has to explicitly compute it. Instead, the value of $\widetilde\Delta (\tilde \gamma_{\log})$ can always be extracted recursively from the action of $\Delta$ on $\tilde \gamma_{\log}$ and its subdivergences \cite{Chetyrkin:1982nn,Smirnov:1986me,Chetyrkin:2017ppe}. The conversion between these two operations can be established from the equation
\beq
R^*\tilde \gamma_{\log} =0\,,
\eeq
which follows immediately from the fact that scaleless vacuum graphs vanish in dimensional regularisation.
Expanding the left hand-side and rearranging we then obtain
\beq
\label{eq:IRtoUV}
\widetilde\Delta (\tilde \gamma_{\log})=-\Delta(\tilde \gamma_{\log})- \sum_{\substack{S\in \bar W_{\emptyset}(\tilde \gamma_{\log}),S'\in \munderbar W_{\emptyset}'(\tilde \gamma_{\log})\\S \cap S' = \emptyset}}\;\widetilde\Delta (S')* \Delta(S) * \tilde \gamma_{\log}/S \setminus S'\,
\eeq
where $\bar W_{\emptyset}$ and $\munderbar W_{\emptyset}$ are proper UV  and IR -woods which exclude the empty graph. The sum over IR and UV  spinneys in the above equation may be simplified 
further by imposing the requirement $S\cup S'=\{\tilde \gamma_{\log}\}$. Terms in the sum not satisfying this requirement would be proportional to scaleless vacuum graphs and hence vanish. 
Below are some simple examples for rewriting IR in terms of UV counterterms.
\bea
\widetilde\Delta \left( \IRonenum{}\right) &=& \widetilde\Delta  \left( \vaconenum{} \right) = -\Delta \left( \vaconenum{} \right)\\
\label{eq:IRCTonetwo}\widetilde\Delta  \left( \IRdotbubblenum{}{}{} \right) &=& \widetilde\Delta  \left( \vactwoloopdotnum{}{}{} \right)= -\Delta \left( \vactwoloopdotnum{}{}{} \right) - \widetilde\Delta \left( \IRonenum{}\right) \Delta \left( \vaconenum{} \right)\nn
\eea

More examples, including IR counterterms with higher SDDs, will be given in the next section.

\subsection{Examples of $R^*$ for generic Feynman graphs}
\label{sec:rstarexample}
The methods which we presented in section \ref{section:Rgeneric} for dealing with arbitrary numerators and tensors within the $R$-operation can be applied in the same manner 
when dealing with IR divergences in the context of the $R^*$-operation. This follows mostly from the fact that the values of IR counterterms can be extracted 
from UV counterterms. However, subtleties arise when IR divergences of higher SDDs are encountered; in particular when they neighbour higher order UV divergences.
In the following we will illustrate how the methods presented above are fully sufficient to tackle all of these cases. 

Below we show an example of a diagram with a linear IR divergence:
\bea
\label{ref:Rstartensorone}
R^* \bubbleonedotdotnum{\alpha}{}{\alpha} &=& \bubbleonedotdotnum{\alpha}{}{\alpha} 
+ \widetilde\Delta  \left ( \IRonedotnum{\alpha} \right) * \arctwonum{}{\alpha}{}\nn\\
 &=& \bubbleonedotdotnum{\alpha}{}{\alpha} 
+ \widetilde\Delta  \left ( \IRonedotnum{\,\alpha \beta} \right) \left[ \partial^\beta_{p_1} \;\arctwonum{}{\alpha}{1} \right]_{p_1=0}\\
 &=& \bubbleonedotdotnum{\alpha}{}{\alpha} 
-2 \widetilde\Delta  \left ( \IRonedotnum{\,\alpha \beta} \right)  \linenumtwo{\alpha}{\beta} \nn
\eea
Here $p_1$ is used to denote the momentum of the IR leg flowing into the remaining diagram. In the second line we evaluated the linear order term of 
the $p_1=0$ Taylor expansion of the remaining graph. We see that this Taylor expansion leaves us with an IR counterterm of a rank $2$ tensor. 
This term can be evaluated using eq.(\ref{eq:IRtoUV}) and the tensor reduction method introduced in section \ref{section:Rtensor}:
\beq
\widetilde\Delta  \left ( \IRonedotnum{\;\alpha \beta} \right)=-\Delta \left ( \bubbleonedotnum{}{\alpha\beta} \right)=g^{\alpha\beta}K\l(\frac{1}{D}\bubbleonenum{}{}\r)\;.
\eeq
Inserting this expression back into eq.(\ref{ref:Rstartensorone}) we then obtain:
\beq
R^* \bubbleonedotdotnum{\alpha}{}{\alpha}=\bubbleonedotdotnum{\alpha}{}{\alpha} +2 K\l(\frac{1}{D}\bubbleonenum{}{}\r) \linenum{}
\eeq
The insertion of higher order UV counterterms and higher order IR counterterms does not commute in general: it is crucial to first insert the UV counterterm and only then to apply the Taylor expansion corresponding to the IR counterterm. This is illustrated in the following example:
\bea
R^* \left( \sunrisedotdot{\mu}{}{\mu} \right) &=& 1*\sunrisedotdot{\mu}{}{\mu} + \Delta \left( \sunrisedotdot{\mu}{}{\mu} \right) * 1 \nn\\ 
&&+ \widetilde\Delta  \left ( \IRonedotnum{\mu} \right) * \bubbleonenumrem{\mu}{}
+\widetilde\Delta  \left ( \IRonedotnum{\mu} \right) 
* \Delta \left( \bubbleonenum{\mu}{} \right) * \vxnum{} \nn\\
	&=& \sunrisedotdot{\mu}{}{\mu} + \Delta \left( \sunrisedotdot{\mu}{}{\mu} \right)
	-2 \widetilde\Delta  \left ( \IRonedotnum{\mu \alpha} \right) \bubbleonedotnum{\mu}{\alpha} \nn\\
	&&-2 \Delta \left( \bubbleonedotnum{\mu}{\alpha} \right) \cdot \left( \widetilde\Delta  \left ( \IRonedotnum{\mu} \right) * 
	 \vxnum{\alpha} \right) \nn\\
	&=& \sunrisedotdot{\mu}{}{\mu} + \Delta \left( \sunrisedotdot{\mu}{}{\mu} \right)
	-2 \widetilde\Delta  \left ( \IRonedotnum{\mu \alpha} \right) \bubbleonedotnum{\mu}{\alpha} \nn\\
	&&-2 \widetilde\Delta  \left ( \IRonedotnum{\mu \beta} \right) \Delta \left( \bubbleonedotnum{\mu}{\alpha} \right) g^{\alpha \beta} \; ,
\eea
where the $p^\alpha$ created by the UV Taylor expansion was acted upon by the IR Taylor expansion.
Substituting the results for the tensor reduced counterterms we then obtain
\begin{align}
\label{eq:iruvexamplequad}
R^* \left( \sunrisedotdot{\mu}{}{\mu} \right) =& \sunrisedotdot{\mu}{}{\mu} + \Delta \left( \sunrisedotdot{\mu}{}{\mu} \right)
- 2 K\l(\frac{1}{D}\bubbleonenum{}{}\r) \bubbleonedotnum{\mu}{\mu} \nn\\
&+ 2 D\, K\l(\frac{1}{D}\bubbleonenum{}{}\r) \,K\l(\frac{1}{D}\bubbleonenum{}{}\r)\,.
\end{align}
The factor of $D$ is produced outside any $K$-operation and signifies the interplay between neighbouring higher order IR and UV  divergences. The fact that the subtraction of IR and UV  divergences does not commute in general was also discussed in \cite{Phi4}. 

We can drastically simplify the computation of the UV counterterm of the above diagram by dropping the external momenta, rewriting the dot
product, and using symmetry:
\bea
\Delta \left( \sunrisedotdot{\mu}{}{\mu} \right) &=& \Delta \left( \vacsunrisedotdot{\mu}{}{\mu} \right) =
\frac{1}{2}\Delta \left( \vacbubbletwodotdot - \vacbubbletwodotdot -  \vacsunrisedot{}{}{} \right) \nn\\ &=& 
-\frac{1}{2}\Delta \left( \vacsunrisedot{}{}{} \right) 
\eea
This shows how LTVGs can be conveniently used to simplify the calculation of counterterms.
Let us now study
\bea
R^* \tonebubbledotnum{\mu}{\mu} &=& \tonebubbledotnum{\mu}{\mu} + \widetilde\Delta \l( \IRonenum{} \r)*\bubbletwodotnumrem{\mu}{\mu} + \Delta \left ( \bubbleonenum{\mu}{} \right) * \bubbletwodotnume{}{}{}{\mu}{}\\
&&+\widetilde\Delta \l( \IRonenum{} \r)*\Delta \left ( \bubbleonenum{\mu}{} \right) * \bubbleonenumremb{}{\mu}+ \widetilde\Delta  \left ( \IRtwonum{}{}{} \right)*\l(\bubbleonenumdis{\mu}{}\arcthreenum{\mu}{}\r)\nn\\
&&+ \widetilde\Delta  \left ( \IRtwonum{}{}{} \right) * \Delta \left ( \bubbleonenumrem{\mu}{} \right) *\arcthreenum{\mu}{}\nn \;.
\label{eq:iruvexample}
\eea
This example shows several interesting features.
First, we consider the three-line IR subgraph which appears in the second and third line.
By constructing its associated contracted vacuum graph we can relate its IR counterterm to one which we already computed in eq.(\ref{eq:IRCTonetwo}):
\beq
\widetilde\Delta  \left ( \IRtwonum{}{}{} \right)= \widetilde\Delta  \left ( \vactwoloopdotnum{}{}{} \right)=\widetilde\Delta  \left( \IRdotbubblenum{}{}{} \right).
\eeq
Second, we notice that this IR subgraph is disjoint in the original graph, a feature which was already discussed for a simpler example in section \ref{IRdivs}. 
As a result, the remaining graph splits into two disjoint components. Since one of the two components has no dependence on external momenta, 
it becomes scaleless when it is acted upon by the Taylor expansion operator, and as a result vanishes:
\beq
\widetilde\Delta  \left ( \IRtwonum{}{}{} \right)*\l(\bubbleonenumdis{\mu}{}\arcthreenum{\mu}{}\r)=0\,.
\eeq
Third, we see that the counterterm in eq.(\ref{eq:iruvexample}) has a linear UV subgraph that will generate a momentum that belongs to the IR in the remaining diagram. In this case, the IR subgraph has logarithmic SDD, 
which means that after Taylor expanding the IR --- setting all IR momenta to 0 --- this counterterm will vanish:
\bea
	&&\widetilde\Delta  \left ( \IRtwonum{}{}{} \right) * \Delta \left ( \bubbleonenum{\mu}{} \right) *
		\arcthreenum{\mu}{}\nn\\
	&=&-2 \Delta \left ( \bubbleonedotnum{\mu}{\alpha} \right) \cdot\l(\widetilde\Delta  \left ( \IRtwonum{}{}{} \right) * \arcthreenum{\mu}{\alpha}\r)\\
	&=& 0 .\nn
\eea
Thus we see that both counterterms  containing this particular IR subgraph vanish, although for completely different reasons.
We continue with another example:
\beq
R^* \budotdotnum{}{}{\mu}{}{}{} = \budotdotnum{}{}{\mu}{}{}{}+\IRonenum{}*\budotdotnumrem{}{}{\mu}{}{}{}+\widetilde\Delta  \left ( \IRtwodotnum{}{}{\mu} \right) * \arcdubblethreenum{}{}{}{} \, ,
\eeq
which shows an IR divergence of linear SDD with three IR legs entering the remaining graph. 
Here one can choose to Taylor expand around any independent set of IR subgraph momenta. 
Taking $p_3=p_{3'}\equiv -p_1-p_2$ we may Taylor expand around $p_1=p_2=0$:
\bea
\widetilde\Delta  \left ( \IRtwodotnum{}{}{\mu} \right) * \arcdubblethreenum{}{}{}{} &=& 
\widetilde\Delta  \left( \IRtwodotnum{\alpha}{}{\mu} \right) \l[ \partial_{p_1}^\alpha \arcdubblethreenum{}{1}{2}{3'}\r]_{p_{1,2}=0} 
+\widetilde\Delta  \left( \IRtwodotnum{}{\alpha}{\mu} \right) \l[ \partial_{p_2}^\alpha \arcdubblethreenum{}{1}{2}{3'}\r]_{p_{1,2}=0} \nn\\
&=&-2 \widetilde\Delta  \left ( \IRtwodotnum{}{\alpha}{\mu} \right) \bubbleonedotdotnum{\alpha}{}{}-2 \widetilde\Delta  \left ( \IRtwodotnum{\alpha}{}{\mu} \right) \bubbleonedotupdownnum{}{\alpha}{} \, .
\eea
The IR counterterms are recursively converted to UV counterterms, i.e.,
\bea
  \widetilde\Delta  \left ( \IRtwodotnum{\alpha}{}{\mu} \right) &=& \widetilde\Delta  \left ( \vactwoloopdotupperlowernum{\mu}{\alpha}{} \right)= -\Delta \left (  \vactwoloopdotupperlowernum{\mu}{\alpha}{} \right) - \widetilde\Delta  \left ( \IRonenum{} \right) \Delta \left ( \bubbleonedotnum{\mu}{\alpha} \right)\nn\\
  \widetilde\Delta  \left ( \IRtwodotnum{}{\alpha}{\mu} \right) &=& \widetilde\Delta  \left ( \vactwoloopdotupperlowernum{\mu}{}{\alpha} \right)= -\Delta \left (  \vactwoloopdotupperlowernum{\mu}{}{\alpha} \right) \, ,
\eea
and then tensor reduced as discussed before.

\subsection{Properties of logarithmic vacuum graphs}
We will now summarise a few properties of counterterms of LTVGs. Even though the UV counterterm operation does not generally commute with contraction, 
it is additive under integrand relations:
\beq
\label{eq:additivity}
\Delta(G_1+G_2)=\Delta(G_1)+\Delta(G_2)\,.
\eeq
In stark contrast, it appears that $\widetilde\Delta $ does not satisfy an analog of the additivity property in eq.(\ref{eq:additivity}). 
Consider for instance the graph:
\beq
\vacsunrisedotdot{\mu}{\mu}{}=\frac{1}{2}\l(\vacsunrisedot{}{}{}-\vacbubbletwodotdot-\vacbubbletwodotdot\r)\,,
\eeq
for which one can easily confirm confirm additivity:
\beq
\label{eq:Deltascalarrewriteex}
\Delta\l(\vacsunrisedotdot{\mu}{\mu}{}\r)=\frac{1}{2}\Delta\l(\vacsunrisedot{}{}{}\r)
,\quad\text{given}\quad\Delta\l(\vacbubbletwodotdot\r)=0\,.
\eeq
However the same is not true for the IR counterterm:
\beq
\widetilde\Delta \l(\vacsunrisedotdot{\mu}{\mu}{}\r)\neq\frac{1}{2}\widetilde\Delta \l(\vacsunrisedot{}{}{}\r)
,\quad\text{given}\quad\widetilde\Delta \l(\vacbubbletwodotdot\r)=0\,.
\eeq
This can be verified by direct computation:
\begin{align}
\widetilde\Delta  \left( \vacsunrisedotdot{\mu}{\mu}{} \right) &= -\Delta \left( \vacsunrisedotdot{\mu}{\mu}{} \right) - 
	\widetilde\Delta  \left( \IRonedotnum{} \right) * \Delta \left( \bubbleonenum{\mu}{\mu} \right) * \vxnum{} \nn\\
	&=-\frac{1}{2} \Delta \left( \vacsunrisedot{}{}{} \right) + \frac{1}{2}D K \left( \frac{1}{D} \bubbleonenum{}{} \right) K \left( \bubbleonenum{}{} \right) \; ,
\end{align}
while instead
\begin{align}
\widetilde\Delta  \left( \vacsunrisedot{}{}{} \right) &= -\Delta \left( \vacsunrisedot{}{}{} \right) - 
	\widetilde\Delta  \left( \IRonenum{} \right) * \Delta \left( \bubbleonenum{}{} \right) * 1 \nn\\
	&=-\Delta \left( \vacsunrisedot{}{}{} \right) + K \left( \bubbleonenum{}{} \right) K \left( \bubbleonenum{}{} \right) \,.
\end{align}
We note that the reason for this apparent disagreement stems from a non-cancellation of $D$s in $K(\frac{1}{D}A)D\neq K(A)$. While a consistent scheme may exist where the IR counterterms could be chosen to be additive, this scheme would likely destroy some of the nice properties of the UV counterterm operation. One could imagine that such a scheme arises naturally if the $R^*$-operation was to be formulated in configuration space, where the roles of IR and UV are effectively interchanged. The absence of additivity of the IR counterterm may appear to be a hindrance in calculations, but it does not present a practical limitation since it is always possible to rewrite IR counterterms in terms of UV counterterms of LTVGs via eq.(\ref{eq:IRtoUV}). In turn, the LTVGs can be simplified using the additivity property of eq.(\ref{eq:additivity}). 

Another useful property of the UV counterterm operation is that it commutes with uncontracted differentiation operators.
This also allows one to derive IBP-like relations: 
\beq
\label{eq:DELTAIBPS}
\Delta(\partial_{p_i}^\mu p_j^\nu \tilde \gamma(\{p_k\}))=0\,, 
\eeq
where $\{p_i\}$ is a set of independent momenta spanning the LTVG $\tilde\gamma$. When applying eq.(\ref{eq:DELTAIBPS}) to LTVGs of a certain tensorial rank,
it returns relations among LTVGs with the tensorial rank raised by up to two. An example is given by the IBP 
\beq
\Delta\l(\partial_{p_1}^{\nu} p_1^\mu\vacsunrisedot{1}{2}{3} \r)=0\,,
\eeq
which yields the relation:
\beq
0=-g_{\mu\nu}\Delta\l(\vacsunrisedot{}{}{}\r)-2\Delta\l(\vacsunrisedotdotb{}{\mu\nu}{}\r)+2\Delta\l(\vacsunrisedotdotb{\mu}{\nu}{}\r)\,.
\eeq
IBP relations thus allow one to find relations between counterterms of LTVGs of different tensorial rank, which can be used to simplify 
calculations. Let us finish this section by giving examples for some common UV counterterms of LTVGs.
We shall use the following normalisation for the integration measure of each independent loop momentum:
\beq
\label{eq:normalisation}
\mu^{2\eps} e^{\eps(\gamma_E+\zeta_2/2)}\int \frac{d^Dk}{\pi^{D/2}} \; ,
\eeq
with $\gamma_E$ the Euler-Mascheroni constant and $\zeta_n$ the Riemann zeta function. 
\bea
&&\Delta\l(\vaconenum{}\r)= \frac{1}{\eps},\qquad \Delta\l(\vaconenumt{\mu_1\mu_2}\r)=\frac{1}{4\eps}g_{\mu_1\mu_2},\nn \\
&&\Delta\bigg(\vaconenumtt{\mu_1\ldots\mu_4}\bigg)=\frac{1}{24\eps} \l(g_{\mu_1\mu_2}g_{\mu_3\mu_4}+g_{\mu_1\mu_3}g_{\mu_2\mu_4} +g_{\mu_1\mu_4}g_{\mu_2\mu_3}\r)\,,\nn\\
&&\Delta\l(\vacsunrisedot{}{}{}\r)=-\frac{1}{2\eps^2}+\frac{1}{2\eps},\qquad \Delta\l(\vacsunrisedotdot{}{}{\mu_1\mu_2}\r)= g_{\mu_1\mu_2} \l(\frac{1}{16\eps}-\frac{1}{8\eps^2}  \r)\,, \\
&&\Delta\l(\vacsunrisedotdotdot{\mu_1\ldots\mu_4}{}{}\r)=-\frac{1}{96\eps} \l(g_{\mu_1\mu_2}g_{\mu_3\mu_4}+g_{\mu_1\mu_3}g_{\mu_2\mu_4} +g_{\mu_1\mu_4}g_{\mu_2\mu_3}\r)\,,\nn\\
&&\Delta\l(\vecthreesunrisea\r)=\frac{1}{3\eps^3}-\frac{2}{3\eps^2}+\frac{1}{3\eps},\qquad \Delta\l(\vacthreebenzdot{\mu_1\mu_2}\r)=g_{\mu_1\mu_2}\frac{\zeta_3}{2\eps}\,.\nn
\eea
\section{Applications of $R^*$}
\label{sec:applications}
In this section we shall demonstrate the $R^*$ method introduced in the previous sections to compute the pole part
of a number of complicated, and so far unknown, five loop propagator integrals. This is achieved through the relations
introduced in section \ref{R}:
\beq
KG=-K \delta R^* G, \qquad  \delta R^* =R^*-1 \, ,
\eeq
which allows one to compute the poles of any $L$-loop propagator integral from propagator integrals of maximally $L-1$ loops, 
a fact which follows straight forwardly from the structure of the $R^*$-operation.


We have implemented the algorithms described in this work in two independent computer codes. 
One is written entirely in \textsc{Form} \cite{Kuipers:2012rf} and the other is mostly written in Maple.
We interfaced both of these implementations with the Forcer program \cite{tuLL2016,tuACAT2016,FORCER}, 
a highly efficient \textsc{Form} program that uses parametric integration-by-parts reduction rules to reduce any 
propagator up to four loops to a set of known master integrals. This makes it possible to compute all five-loop propagator integrals from the knowledge of up to four-loop integrals. The combination of the $R^*$ algorithm presented in this work and the Forcer program was used in the recent computation of the five-loop beta function in QCD in a general gauge group \cite{Herzog:2017ohr}, which took three days on a cluster. It is worth emphasising that our application of the $R^*$-operation differs in spirit not only from the global approach of \cite{Baikov:2016tgj}, but also from the local approach which was taken for instance in \cite{Gorishnii:1983gp,Kleinert:1991rg,Kompaniets:2016hct} to compute the six loop anomalous dimensions in scalar $\phi^4$ theory. Indeed, the local $R^*$ approach taken there was based on the application of $K\bar R^*$, which allows one to directly isolate the renormalisation group constant of the relevant order. 
Our approach focuses on simply computing the poles of a given amplitude from which the corresponding renormalisation group constants can of course be extracted. In the case of non-scalar QFTs, such as QCD, there are advantages in our approach. While the UV counterterm operation $K\bar R^*$ is sensitive to the contraction anomalies, discussed extensively in section \ref{section:Rgeneric}, the operation $K\delta R^*$ is insensitive to them. 
This observation allows us to make full use of integrand (and even integral) relations of generic Feynman graphs. In contrast the traditional local approach with $K\bar R^*$ would not easily allow for the use of such relations.

We shall start by presenting results for the poles of a number of five-loop integrals in $\phi^3$-theory in four dimensions. While these integrals are in fact not superficially UV divergent, they do contain highly intricate 
IR divergences with high SDDs.
Furthermore, some of these integrals are part of a yet to be found basis of five-loop master integrals, and we thus 
anticipate that the explicit results for their poles will constitute useful benchmarks for future 
evaluations of their finite parts. All results in the following will be normalised according 
to eq.(\ref{eq:normalisation}), and we set $Q^2=\mu^2=1$, where $Q$ is the (Euclidean) external momentum.

Below we present the pole part of five $\phi^3$ topologies:
\begin{align}
K
\raisebox{-21pt}
{
\begin{axopicture}{(78,50)(-36,-24)}
\SetScale{1}\SetColor{Black}%
\CCirc(0,0){20}{Black}{White}
\Line(-30,0)(-20,0)
\Line(20,0)(30,0)
\Line(-14.14,14.14)(14.14,-14.14)
\Line(-14.14,-14.14)(14.14,14.14)
\Line(-14.14,-14.14)(14.14,14.14)
\Line(-10,-10)(-10,10)
\CCirc(0,0){5}{White}{White}
\Line(0,-20)(0,20)
\end{axopicture}
}
=& -\frac{56}{\eps} \zeta_7 \\
K
\raisebox{-21pt}
{
\begin{axopicture}{(78,50)(-36,-24)}
\SetScale{1}\SetColor{Black}%
\CCirc(0,0){20}{Black}{White}
\Line(-30,0)(-20,0)
\Line(20,0)(30,0)
\Line(-14.14,14.14)(0,-10)
\Line(14.14,14.14)(0,-10)
\CCirc(-5,0){4}{White}{White}
\CCirc(5,0){4}{White}{White}
\Line(-14.14,-14.14)(0,10)
\Line(14.14,-14.14)(0,10)
\Line(0,-20)(0,-10)
\Line(0,20)(0,10)
\end{axopicture}
}
=& \frac{4}{\eps^2} \zeta_5 + \frac{1}{\eps} \left( 10 \zeta_6 + 4 \zeta_3^2 \right)\\
K
\raisebox{-21pt}
{
\begin{axopicture}{(78,50)(-36,-24)}
\SetScale{1}\SetColor{Black}%
\CCirc(0,0){20}{Black}{White}
\Line(-30,0)(-20,0)
\Line(20,0)(30,0)
\Line(-19.5,5)(19.5,5)
\CCirc(0,5){4}{White}{White}
\Line(-14.14,14.14)(0,10)
\Line(14.14,14.14)(0,10)
\Line(0,-10)(0,10)
\Line(-14.14,-14.14)(0,-10)
\Line(14.14,-14.14)(0,-10)
\end{axopicture}
}
=& -\frac{9}{10 \eps^3} \zeta_3 + \frac{1}{\eps^2} \left(- \zeta_5 - \frac{27}{20} \zeta_4 - \frac{81}{20} \zeta_3 \right)\\
   &+\frac{1}{\eps} \left(-\frac{5}{2} \zeta_6
+\frac{1}{5} \zeta_5 -\frac{243}{40} \zeta_4 + \frac{159}{8} \zeta_3
   -5 \zeta_3^2 \right) \nn\\
K
\raisebox{-21pt}
{
\begin{axopicture}{(78,50)(-36,-24)}
\SetScale{1}\SetColor{Black}%
\CCirc(0,0){20}{Black}{White}
\Line(-30,0)(-20,0)
\Line(20,0)(30,0)
\Line(-19.5,5)(19.5,5)
\CCirc(-5,5){4}{White}{White}
\CCirc(3,5){4}{White}{White}
\Line(-14.14,14.14)(0,0)
\Line(0,-20)(0,0)
\Line(0,0)(0,20)
\Line(0,10)(14.14,14.14)
\end{axopicture}
}
=& 
       \frac{1}{5 \eps^3} \zeta_3 + \frac{1}{\eps^2} \left( \frac{3}{10} \zeta_4 + \frac{1}{2} \zeta_3 \right)\\
         &+\frac{1}{\eps} \left( - \frac{147}{40} \zeta_7 + \frac{11}{15} \zeta_5 + \frac{3}{4} \zeta_4
          - \frac{641}{60} \zeta_3 + \frac{1}{5} \zeta_3^2 \right) \nn\\
K
\raisebox{-21pt}
{
\begin{axopicture}{(78,50)(-36,-24)}
\SetScale{1}\SetColor{Black}%
\CCirc(0,0){20}{Black}{White}
\Line(-30,0)(-20,0)
\Line(20,0)(30,0)
\Line(17.32,10)(-17.32,-10)
\Line(-17.32,10)(17.32,-10)
\Line(-10,17.32)(10,-17.32)
\CCirc(0,0){2}{White}{White}
\Line(10,17.32)(-10,-17.32)
\end{axopicture}
}=&  \frac{6}{\eps^2} \zeta_5 + \frac{1}{\eps}\left( 42 \zeta_7 + 15 \zeta_6 + 21 \zeta_5 + 6 \zeta_3^2 \right)
\end{align}
The computation of the pole parts only takes a few seconds on a single core. 
A list of all top-level five-loop $\phi^3$ topologies will be provided
on \texttt{arXiv.org} as an attachment to this article.

Next, we present the pole parts of some five-loop ghost propagator diagrams, where we
enforce that the ghost line goes through all vertices. If we use the Feynman gauge,
the QCD diagram has a one-to-one correspondence to a generic Feynman graph.
This is illustrated below:
\begin{align}
\raisebox{-21pt}
{
\begin{axopicture}{(60,50)(-30,-24)}
\SetScale{1}\SetColor{Black}%
\SetArrowScale{0.6}
\GluonArc(0,0)(20,0,45){2}{2}
\Arc[dash,dsize=2,arrow,flip](0,0)(20,45,90)
\GluonArc(0,0)(20,90,135){2}{2}
\Arc[dash,dsize=2,arrow](0,0)(20,135,180)
\GluonArc(0,0)(20,180,225){2}{2}
\Arc[dash,dsize=2,arrow,flip](0,0)(20,225,270)
\Arc[dash,dsize=2,arrow,flip](0,0)(20,270,315)
\Arc[dash,dsize=2,arrow,flip](0,0)(20,315,360)
\Line[dash,dsize=2,arrow,flip](-30,0)(-20,0)
\Line[dash,dsize=2,arrow,flip](20,0)(30,0)
\Line[dash,dsize=2,arrow,flip,arrowpos=0.25](-14.14,14.14)(0,-10)
\Line[dash,dsize=2,arrow,arrowpos=0.25](14.14,14.14)(0,-10)
\CCirc(-5,0){3}{White}{White}
\CCirc(5,0){3}{White}{White}
\Line[dash,dsize=2,arrow,arrowpos=0.25](-14.14,-14.14)(0,10)
\Gluon(14.14,-14.14)(0,10){2}{4}
\Gluon[arrow](0,-20)(0,-10){2}{2}
\Line[dash,dsize=2,arrow,flip](0,20)(0,10)
\Vertex(-20,0){1}
\Vertex(20,0){1}
\Vertex(0,-20){1}
\Vertex(0,20){1}
\Vertex(14.14,14.14){1}
\Vertex(-14.14,14.14){1}
\Vertex(14.14,-14.14){1}
\Vertex(0,10){1}
\Vertex(0,-10){1}
\end{axopicture}
}
\rightarrow
\raisebox{-21pt}
{
\begin{axopicture}{(60,50)(-30,-24)}
\SetScale{1}\SetColor{Black}%
\SetArrowScale{0.6}
\Arc(0,0)(20,0,45)
\Arc[arrow,flip](0,0)(20,45,90)
\Arc(0,0)(20,90,135)
\Arc[arrow](0,0)(20,135,180)
\Arc(0,0)(20,180,225)
\Arc[arrow,flip](0,0)(20,225,270)
\Arc[arrow,flip](0,0)(20,270,315)
\Arc[arrow,flip](0,0)(20,315,360)
\Line[arrow,flip](-30,0)(-20,0)
\Line[arrow,flip](20,0)(30,0)
\Line[arrow,flip,arrowpos=0.25](-14.14,14.14)(0,-10)
\Line[arrow,arrowpos=0.25](14.14,14.14)(0,-10)
\CCirc(-5,0){3}{White}{White}
\CCirc(5,0){3}{White}{White}
\Line[arrow,arrowpos=0.25](-14.14,-14.14)(0,10)
\Line(14.14,-14.14)(0,10)
\Line(0,-20)(0,-10)
\Line[arrow,flip](0,20)(0,10)
\Text(-25,-5) {\tiny $\mu$ \tiny}
\Text(-21,11) {\tiny $\rho$ \tiny}
\Text(-12,5) {\tiny $\kappa$ \tiny}
\Text(-12,-5) {\tiny $\mu$ \tiny}
\Text(-5,-15) {\tiny $\rho$ \tiny}
\Text(8,-15) {\tiny $\sigma$ \tiny}
\Text(16,-6) {\tiny $\nu$ \tiny}
\Text(14,5) {\tiny $\nu$ \tiny}
\Text(-3,14) {\tiny $\sigma$ \tiny}
\Text(8,14) {\tiny $\rho$ \tiny}
\end{axopicture}
}
=&
   \frac{1}{\eps^2}\left(\frac{11}{2560} -\frac{1}{64} \zeta_5 +\frac{3}{256} \zeta_3 \right)
    +\frac{1}{\eps}\bigg(\frac{551}{5120}-\frac{5}{128} \zeta_6 \nn\\
    &-\frac{109}{256} \zeta_5
         +\frac{9}{512} \zeta_4
         +\frac{729}{2560} \zeta_3
         +\frac{1}{32} \zeta_3^2
         \bigg)
         + \mathcal{O}(\eps^0)
\end{align}
Every Feynman diagram of this type has five dot products. As a result, there will be many
tensor UV subgraphs. We have not rewritten the dot products to a basis, since this will create
higher-order UV and IR divergences. Any speed gains from the simplified topologies are negated by
expensive Taylor expansions and tensor reductions. 

Below we present three more examples:
\begin{align}
K
\raisebox{-21pt}
{
\begin{axopicture}{(60,50)(-30,-24)}
\SetScale{1}\SetColor{Black}%
\SetArrowScale{0.6}
\Arc[arrow](0,0)(20,0,45)
\Arc[arrow](0,0)(20,45,90)
\Arc[arrow](0,0)(20,90,135)
\Arc[arrow](0,0)(20,135,180)
\Arc[](0,0)(20,180,225)
\Arc[arrow,flip](0,0)(20,225,270)
\Arc[arrow,flip](0,0)(20,270,315)
\Arc[](0,0)(20,315,360)
\Line[arrow,flip](-30,0)(-20,0)
\Line[arrow,flip](20,0)(30,0)
\Line(14.14,14.14)(14.14,0)
\Line[arrow](14.14,0)(14.14,-14.14)
\Line(-14.14,14.14)(-14.14,0)
\Line[arrow,flip](-14.14,0)(-14.14,-14.14)
\Line(0,20)(0,-20)
\CCirc(0,0){3}{White}{White}
\Line[arrow](-14.14,0)(14.14,0)
\Text(-25,-5) {\tiny $\mu$ \tiny}
\Text(-21,9) {\tiny $\rho$ \tiny}
\Text(-7,23) {\tiny $\kappa$ \tiny}
\Text(10,23) {\tiny $\sigma$ \tiny}
\Text(23,9) {\tiny $\nu$ \tiny}
\Text(-7,-23) {\tiny $\kappa$ \tiny}
\Text(10,-23) {\tiny $\nu$ \tiny}
\Text(-8,-8) {\tiny $\mu$ \tiny}
\Text(10,-8) {\tiny $\sigma$ \tiny}
\Text(-5,3) {\tiny $\rho$ \tiny}
\end{axopicture}
}
=& 
  \frac{1}{\eps^2}\left(
         \frac{1}{512}
         -\frac{1}{320} \zeta_3
         \right) +\\&
    \frac{1}{\eps}\left(
         \frac{337}{5120}
         -\frac{161}{1280}\zeta_7
         -\frac{25}{128}\zeta_5
         -\frac{3}{640}\zeta_4
         +\frac{341}{1280}\zeta_3
         -\frac{9}{160}\zeta_3^2
         \right)\nn
 \\
K
\raisebox{-21pt}
{
\begin{axopicture}{(60,50)(-30,-24)}
\SetScale{1}\SetColor{Black}%
\SetArrowScale{0.6}
\Arc[](0,0)(20,0,60)
\Arc[arrow](0,0)(20,60,90)
\Arc[arrow](0,0)(20,90,120)
\Arc[](0,0)(20,120,150)
\Arc[arrow](0,0)(20,150,180)
\Arc[](0,0)(20,180,240)
\Arc[arrow](0,0)(20,240,300)
\Arc[](0,0)(20,300,330)
\Arc[arrow,flip](0,0)(20,330,360)
\Line[arrow,flip](-30,0)(-20,0)
\Line[arrow,flip](20,0)(30,0)
\Line[arrow,flip,arrowpos=0.25](10,17.32)(10,-17.32)
\Line[arrow,arrowpos=0.75](-10,17.32)(-10,-17.32)
\CCirc(-10,6){3}{White}{White}
\CCirc(10,-6){3}{White}{White}
\Line[arrow,flip,arrowpos=0.75](-17.32,10)(0,0)
\Line[arrow,flip,arrowpos=0.25](0,0)(17.32,-10)
\Line[](0,0)(0,20)
\Text(-25,-5) {\tiny $\mu$ \tiny}
\Text(-22,7) {\tiny $\sigma$ \tiny}
\Text(-7,23) {\tiny $\kappa$ \tiny}
\Text(-3,6) {\tiny $\kappa$ \tiny}
\Text(10,23) {\tiny $\nu$ \tiny}
\Text(23,-7) {\tiny $\nu$ \tiny}
\Text(1,-24) {\tiny $\mu$ \tiny}
\Text(-5,-8) {\tiny $\sigma$ \tiny}
\Text(8,8) {\tiny $\rho$ \tiny}
\Text(6,-8) {\tiny $\rho$ \tiny}
\end{axopicture}
}
=& \frac{1}{\eps^2}\left(
         \frac{1}{2560}
         +\frac{1}{64}\zeta_5
         -\frac{3}{128}\zeta_3
         \right) + \\
    &\frac{1}{\eps}\left(
         -\frac{17}{5120}
         +\frac{161}{640}\zeta_7
         +\frac{5}{128}\zeta_6
         +\frac{19}{128}\zeta_5
         -\frac{9}{256}\zeta_4
         -\frac{1291}{2560}\zeta_3
         +\frac{1}{40}\zeta_3^2
         \right) \nn\\
K
\raisebox{-21pt}
{
\begin{axopicture}{(60,50)(-30,-24)}
\SetScale{1}\SetColor{Black}%
\SetArrowScale{0.6}
\Line[arrow,flip](-30,0)(-20,0)
\Line[arrow,flip](20,0)(30,0)
\Arc[arrow](0,0)(20,0,36)
\Arc[](0,0)(20,36,72)
\Arc[arrow](0,0)(20,72,108)
\Arc[](0,0)(20,108,144)
\Arc[arrow](0,0)(20,144,180)
\Arc[](0,0)(20,180,216)
\Arc[arrow,flip](0,0)(20,216,252)
\Arc[](0,0)(20,252,288)
\Arc[arrow,flip](0,0)(20,288,324)
\Arc[](0,0)(20,324,360)
\Line[arrow](16.18,11.756)(16.18,-11.756)
\Line[arrow,flip](-16.18,11.756)(-16.18,-11.756)
\Line[arrow,flip,arrowpos=0.75](6.18,19.02)(6.18,-19.02)
\Line[arrow,arrowpos=0.25](-6.18,19.02)(-6.18,-19.02)
\Text(-25,-5) {\tiny $\mu$ \tiny}
\Text(-22,7) {\tiny $\nu$ \tiny}
\Text(2,23) {\tiny $\sigma$ \tiny}
\Text(-12,-21) {\tiny $\rho$ \tiny}
\Text(14,-21) {\tiny $\kappa$ \tiny}
\Text(-10,0) {\tiny $\mu$ \tiny}
\Text(0,10) {\tiny $\nu$ \tiny}
\Text(4,-10) {\tiny $\rho$ \tiny}
\Text(13,0) {\tiny $\sigma$ \tiny}
\Text(25,6) {\tiny $\kappa$ \tiny}
\end{axopicture}
}
=& \frac{1}{\eps^2}\left(
         -\frac{7}{5120}
         -\frac{1}{128}\zeta_5
         +\frac{1}{128}\zeta_3
         \right)+ \\
    &\frac{1}{\eps}\left(
         \frac{73}{15360}
         +\frac{441}{2560}\zeta_7
         -\frac{5}{256}\zeta_6
         -\frac{29}{64}\zeta_5
         +\frac{3}{256}\zeta_4
         +\frac{951}{2560}\zeta_3
         -\frac{71}{640}\zeta_3^2
         \right) \nn
\end{align}

Finally, we present an example of a hard four-loop diagram:
\begin{align}
	K \no1example =& -\frac{13}{2304 \eps^4} + \frac{1789}{55296 \eps^3} + \frac{91757}{331776 \eps^2} 
	- \frac{17}{256 \eps^2} \zeta_3 \nn\\ 
	&+ \frac{1}{\eps}\left(\frac{199997}{248832} + \frac{5}{4} \zeta_5 
	- \frac{51}{512} \zeta_4 - \frac{17797}{13824} \zeta_3 \right)
\end{align}
This diagram has
 six dot products, one quadratic IR line, two quadratic UVs, and several logarithmic UV and IR subgraphs.
 The diagram requires 187 unique, scalarised counterterms to compute the diagram. 
 It is interesting to compare the time required for computing the poles of this Feynman graph with 
 the $R^*$-method with the time which a direct computation takes using the Forcer program.  
 On a single core it takes 150 seconds to obtain the result with the $R^*$-operation. 
 A direct computation with the Forcer program takes 675 seconds, 4.5 times as long. 
 The reason why a direct computation is slower is because this particular integral requires reductions
 of seven master topologies at four loops, which are generally slow. 
 In contrast, the $R^*$-operation only requires counterterms of up to three loops 
 and thus avoids these complex reductions. This shows that sometimes it is beneficial to use
 $R^*$, even if a direct reduction is available.

\section{Discussion of the literature}
\label{sec:discussion}

In this section we are going to discuss differences of the $R^*$-method proposed in this work with those of the literature. 

\begin{itemize}
\item \textbf{d'Alembertian versus Taylor expansion}
 
In \cite{Phi4} it is proposed that the UV counterterm of a single scale quadratic integral can
be computed from logarithmic ones by taking the d'Alembertian:
\beq
\Delta(G) = Q^2 \Delta \left(\frac{\Box}{2D} G\right),\qquad \Box = g^{\mu \nu} \frac{\partial}{\partial Q^\mu} \frac{\partial}{\partial Q^\nu} \, .
\eeq
To evaluate the d'Alembertian acting on a Feynman graph, one must choose a path which routes the flow of the external 
momentum through the graph. Whereas the value of the integral $\Box G$ should be independent of the path taken, this is not necessarily 
true for the counterterm operation $\Delta(\frac{\Box}{2D} G)$ which as we discussed in section \ref{section:Rgeneric} is not generally 
invariant under contractions with $g_{\mu\nu}$. Below we give an example where this problem becomes apparent:
\begin{align}
\Delta \l( \sunrisebubble{}{}{} \r) &= -\Delta \l( \sunrise{}{}{} \r) \Delta \l( \bubbleonenum{}{}{} \r) = \frac{1}{4 \eps^2} \\
\Delta \l( \frac{\Box}{2D} \sunrisebubble{}{}{} \r) &= \frac{1}{4 \eps^2} - \frac{1}{8 \eps}
\end{align}
We can easily see from the cutvertex rule that the UV counterterm cannot have a $\frac{1}{\eps}$ pole.
In most $\phi^4$ topologies one may get the correct result by using the d'Alembertian, but
this is in no way guaranteed. Consider for example the following diagram:
\beq
\sunriseselfenergy{}{}{}
\eeq
If one chooses to evaluate the d'Alembertian along a path through the sunrise subgraph on the top, 
a wrong answer is obtained. Instead, the only safe procedure for computing the UV counterterm of higher order UV or IR subgraphs, 
is to perform a Taylor expansion. That is, instead of the d'Alembertian the following differential operator should be used:
\beq
\frac{1}{2}Q_\mu Q_\nu \partial^\mu \partial^\nu\,, 
\eeq
which is guaranteed to commute with the counterterm operation. 

\item \textbf{IR counterterm operation}

The IR counterterm operation proposed in \cite{Chetyrkin:1982nn,Chetyrkin:1984xa,Smirnov:1986me,Phi4,Larin:2002sc,Batkovich:2014rka} is usually formulated as 
\beq
\label{eq:IRCT}
\widetilde\Delta (\gamma')=P_{\gamma'}(\{\partial_{p_i}\})\prod_{k=1}^{l(\gamma')} (2\pi)^D \mu^{-2\eps} \delta^{(D)}(p_i)\;.
\eeq
Here $l(\gamma')$ is the number of loops of the IR subgraph $\gamma'$, $P_{\gamma'}(\{\partial_{p_i}\})$ is a 
homogeneous polynomial of degree $\tilde\omega(\gamma')$ in the differential operators $\{\partial_{p_1},\ldots,\partial_{p_l}\}$ 
and $\{p_1,\ldots,p_{l(\gamma')}\}$ is a set of independent momenta spanning the IR subgraph. 
The strategy to compute the IR counterterm in \cite{Chetyrkin:1982nn,Chetyrkin:1984xa,Smirnov:1986me,Phi4,Larin:2002sc,Batkovich:2014rka} relies on setting up a system of equations, 
by inserting the IR subgraph $\gamma'$ into several suitable graphs, and demanding the coefficients of $P_\gamma'(\{\partial_{p_i}\})$ to render this system finite 
after acting with $R^*$. It is straightforward to show that our 
definition of the IR counterterm given in eq.(\ref{eq:deltaprime}) leads to 
a similar form to that of eq.(\ref{eq:IRCT}). The main difference in our approach is that we directly compute the coefficients of 
$P_\gamma'(\{\partial_{p_i}\})$ from the values of tensor IR subgraphs of logarithmic SDD. The advantage of 
\cite{Chetyrkin:1982nn,Chetyrkin:1984xa,Smirnov:1986me,Phi4,Larin:2002sc,Batkovich:2014rka} 
is that no extra tensor reduction due to the Taylor expansion has to be considered. Even though no such examples exist in the literature, 
this approach will require tensor reduction as well when applied to generic Feynman graphs. Since our entire setting relies on reducing both UV and 
IR counterterms to a common basis of LTVGs this is a small price to pay and allows a unified setting for 
the computation of the UV and IR counterterms, whose linear dependence is neatly expressed through eq.(\ref{eq:IRtoUV}).

\item \textbf{Factorisation of the $R^*$-operation}

Another point we wish to raise, concerns the factorisation of the $R^*$-operation into a pure UV subtraction $R$-operation 
and a pure IR subtraction operation $\tilde R$, as noted in e.g. \cite{Larin:2002sc,Chetyrkin:2017ppe}:
\beq
\label{eq:R*factorisation}
R^*=\tilde R\,R\,.
\eeq
It has already been shown in \cite{Larin:2002sc,Phi4} that for local $R^*$, higher degree IR divergences the $\tilde R$ and $R$ do not generally commute:
\beq
R^* \neq R\,\tilde R.
\eeq
We wish to point out that even $R^*=\tilde R\,R\,$ cannot be naively applied to generic Feynman graphs and is explicitly broken 
in graphs where higher degree IR subgraphs neighbour higher degree UV subgraphs. 
To illustrate this, we use the same example as in \cite{Phi4} where it was used to show the non-commutativity of the UV and IR counterterm operators. Using the relation 
\beq
\Delta \left(\sunrise{}{}{}\right) = - K \left(\frac{1}{Q^2}\sunrise{}{}{}\right) Q^2,
\eeq
which is valid only because the subdivergences of this Feynman graph are vanishing, we get:
\begin{align}
\label{eq:unfactorisedRstar}
K\bar R^* &\left( \sunrisethreeloopdotdot{}{}{} \right) \nn\\ 
&=K\l[\sunrisethreeloopdotdot{}{}{}  + \widetilde\Delta  \left( \IRonedotnum{} \right) * \remsunriseir{}{}{} + \widetilde\Delta  \left( \IRonedotnum{} \right)* \Delta \left( \sunrise{}{}{} \right) * \vxnum{}  \r] \\
&= K\l[\sunrisethreeloopdotdot{}{}{} + \widetilde\Delta  \left( \IRonedotnum{} \right) * \remsunriseir{}{}{} + D K \left( \frac{1}{D} \bubbleonenum{}{} \right) K \left(\frac{1}{Q^2} \sunrise{}{}{} \right) \r]\; .\nn
\end{align}
In contrast, the factorised approach yields
\begin{align}
\label{eq:factorisedRstar}
K \tilde R\,&\bar R\, \left( \sunrisethreeloopdotdot{}{}{} \right)\nn\\
&= K\tilde R \Bigg[ \sunrisethreeloopdotdot{}{}{} - K \left(\frac{1}{Q^2} \sunrise{}{}{} \right) \left(Q^2 \tadonedotdotnum{}{} + 2 \tadonedotdotnum{\mu}{\mu} + \tadonedotnum{}{}\right) \Bigg] \\
&=K\l[ \sunrisethreeloopdotdot{}{}{} + \widetilde\Delta  \left( \IRonedotnum{} \right) * \remsunriseir{}{}{} + K \left( \bubbleonenum{}{} \right)K \left(\frac{1}{Q^2} \sunrise{}{}{} \right)\r] \; .\nn
\end{align}
Thus we see that different results are obtained using the factorised or non-factorised approach. By IR rearrangement,
we can rewrite the counterterm to be IR finite:
\beq
\label{eq:unfactorisedRstarIRR}
K\bar R \left( \sunrisethreeloopdotdotb{}{}{} \right) = K \left[ \sunrisethreeloopdotdotb{}{}{}	 
- K \left(\frac{1}{Q^2} \sunrise{}{}{} \right) \bubbleonenum{}{} \right]  \, .
\eeq
The result of eq.(\ref{eq:unfactorisedRstarIRR}) agrees with eq.(\ref{eq:unfactorisedRstar}), as was shown in \cite{Phi4} as well, and thereby clearly falsifies the approach taken in eq.(\ref{eq:factorisedRstar}). This example further illustrates the importance of factors of $D$ which are created in the interplay of higher degree IR and UV divergences. 

%


\end{itemize}

\section{Conclusions and outlook}
\label{sec:conclusion}
The $R^*$-operation is a powerful tool to compute the poles of arbitrary euclidean Feynman graphs 
with non-exceptional external momenta from simpler Feynman graphs. 
In this work we have extended the $R^*$-operation to Feynman graphs with arbitrary numerators and of 
arbitrary tensorial rank. Since the local $R^*$-operation had previously only been applied to 
scalar theories, we have vastly generalised its range of applicability. 

The methods proposed in this work make full use of rewriting the counterterm 
operations for arbitrary divergent UV  and IR subgraphs in terms of scaleless tensor vacuum graphs 
of logarithmic superficial degree of divergence, which we called LTVGs. This concept, which to the best of 
our knowledge has not previously been employed to this extent, allows one to take advantage of the enhanced symmetry properties of vacuum graphs. 
We analysed contraction anomalies, which are easily traceable within dimensional regularisation,
and provided a consistent scheme that uses LTVGs as basic building blocks for UV and IR counterterms.
Additionally, we have refined the definition and evaluation of the IR counterterm operation, so that it resembles
its UV counterpart. 

These methods have been implemented in efficient computer code and have already been put to the test 
in the evaluation of the five loop beta function of QCD with an arbitrary simple gauge group 
\cite{Herzog:2017ohr}. As a further proof of concept we provided results for the poles of
all five-loop top-level propagator graphs in $\phi^3$ theory in four dimensions, 
as well as several explicit results for five-loop ghost propagator graphs with highly non-trivial 
numerator structures. We envisage that these results will provide useful cross-checks once analytic computations, 
which also contain the finite parts of these Feynman graphs, become available. We are planning to publish 
our implementation of this algorithm in a future publication. 

Besides providing an efficient computational tool, our $R^*$ method may shed some further light on 
an old puzzle related to the absence of certain higher zeta values in the anomalous dimensions, 
such as the beta-function in QCD. While some explanations for this phenomenon have been given in 
\cite{Baikov:2010hf}, we believe that the many relations among LTVGs should allow to 
further illuminate the origin of the absence of these zeta-values, 
as they can clearly be traced to the UV counterterms of only a handful of LTVGs.

The method presented in this paper, in combination with the Forcer program that can compute four-loop
massless propagator diagrams, can be used to compute the poles of any propagator diagram up to five loops.
Among many other applications, these techniques could be used to compute Mellin moments of structure functions 
at five loops, a problem which is currently well out of reach with any other method known to us.

\section{Acknowledgements}

We would like to thank Jos Vermaseren and Andreas Vogt for motivating us to work on this problem and 
useful comments on the paper, Konstantin Chetyrkin for sharing with us many deep insights about the $R^*$-operation
and providing valuable comments on the paper, Erik Panzer for valuable discussions 
on IR divergences and motic subgraphs, Takahiro Ueda for collaboration on the optimisation of the FORM implementation,
and Giulio Falcioni and Eric Laenen for comments on the paper. This work is supported by the 
ERC Advanced Grant no. 320651, ``HEPGAME''.

\glsaddall
\printglossaries

\appendix 

\section{Glossary}
\label{sec:glossary}

Below we give an overview of commonly used abbreviations in this paper.

\begin{description}
\item[IBP] Integration By Parts
\item[IR] Infrared
\item[IRI] Infrared Irreducible
\item[IRR] Infrared Rearrangement
\item[LTVG] Logarithmic Tensor Vacuum Graph
\item[MS] Minimal Subtraction
\item[SDD] Superficial Degree of Divergence
\item[UV] Ultraviolet
\end{description}

\section{Cutvertex rule for scalar diagrams}
\label{Appendix:Cutvertexrule}
The cutvertex rule states that
\begin{equation}
\Delta(\gamma_1\gamma_2)=\Delta(\gamma_1)\Delta(\gamma_2) \, .
\end{equation}
This statement can be proven by induction. We start by proving that the statement 
holds true for the trivial case, where both $\gamma_1$ and $\gamma_2$ contain no subdivergences.
This can be proven as follows:
\begin{align}
\begin{split}
\Delta(\gamma_1\gamma_2)&=-K\bar R( \gamma_1\gamma_2)\\
&=-K\big(\g_1\g_2+\D(\g_1)\g_2+\D(\g_2)\g_1\big)\\
&=-K\big((\g_1+\D(\g_1))(\g_2+\D(\g_2))-\D(\g_1)\D(\g_2)\big)\\
&=-K\big(R(\g_1)R(\g_2)-\D(\g_1)\D(\g_2)\big)\\
&=K\big(\D(\g_1)\D(\g_2)\big)\\
&=\D(\g_1)\D(\g_2) \;.
\end{split}
\end{align}
Now we can prove inductively that the same holds for the general case, where we assume that both $\gamma_1$ and $\gamma_2$
have subdivergences. That is, we show that 
\begin{equation}
\D(G_1G_2)=\D(G_1)\D(G_2)  
\end{equation}
holds, assuming the induction hypothesis $\D(\g_1\g_2)=\D(\g_1)\D(\g_2)$ where $\gamma_1$ and $\gamma_2$ 
are subgraphs of $G_1$ and $G_2$ respectively. Let us start with the definition:
\begin{eqnarray}
\Delta(G_1G_2)&=&-K\bar R( G_1G_2)\nn\\
&=&-K\sum_{S\in \bar W(G_1G_2)} \D(S)*G_1G_2/S\,.
\end{eqnarray}
We will now use the fact that we can write
\begin{equation}
\bar W(G_1G_2)=W(G_1)\times W(G_2)\setminus \{\{G_1\},\{G_2\} \}
\end{equation}
with $\times$ denoting the Cartesian product of two sets. This in turn implies
\begin{eqnarray}
\Delta(G_1G_2)&=&-K\bigg[ \sum_{S_1\in W(G_1)}\sum_{S_2\in W(G_2)} \D(S_1 S_2)*G_1G_2/S_1/S_2\  - \D(G_1)\D(G_2)   \bigg]\nonumber\\
\end{eqnarray}
Assuming the induction hypothesis $\D(S_1 S_2)=\D(S_1)\D(S_2)$ we then get
\begin{eqnarray}
\Delta(G_1G_2)&=&-K\bigg[R(G_1)R(G_2)  - \D(G_1)\D(G_2)   \bigg] \nn\\
&=& \D(G_1)\D(G_2) \, .
\end{eqnarray}

\section{Cutvertex rule for contracted tensor diagrams}
\label{Appendix:CutvertexruleTensor}
If weakly non-overlapping (no common edges) subgraphs $\g_1$ and $\g_2$ contain contracted Lorentz indices,
one has in general
\begin{equation}
K\big(\D(\g_1)\D(\g_2)\big)\neq\D(\g_1)\D(\g_2)\;.
\end{equation}
This means that the proof for the factorisation of the counterterm operation $\D$ given in appendix \ref{Appendix:Cutvertexrule} 
breaks down. As a result, it is rather difficult to derive a corresponding generalised ``cut-vertex rule'' for the case of 
contracted tensor subgraphs that does not result in a change of renormalisation scheme. However, when one is interested only in computing the poles of a factorised Feynman graph $G_1G_2$ 
via the use of the identity
\begin{equation}
KG=-K\,\delta R G \, ,
\end{equation}
we will show that the following cutvertex rule still holds:
\begin{equation}
\label{eq:fac}
\D(G_1G_2)\to\D(G_1)\D(G_2) \, .     
\end{equation}
We can actually prove this statement rather easily by noting that the $R$-operation computed with eq.~(\ref{eq:fac})
results in the following replacement:
\begin{equation}
R(G_1G_2)\to R(G_1)R(G_2)  \, .
\end{equation}
We can now write
\begin{equation}
\delta R (G_1G_2)=R(G_1G_2)-G_1G_2=R(G_1)R(G_2)-G_1G_2+\xi \;,
\end{equation}
where $\xi$ denotes the ``error'' one makes by computing with eq.~(\ref{eq:fac}).
From this it follows that
\begin{equation}
\xi=   R(G_1G_2)-R(G_1)R(G_2)\, .
\end{equation}
Given that $\xi$ is manifestly finite, we obtain:
\begin{equation}
K\xi=0\Rightarrow   K\delta R (G_1G_2)=K R(G_1)R(G_2)-K G_1G_2\;.
\end{equation}
This completes the proof that the poles of a factorised graph can be computed by consistently applying eq.(\ref{eq:fac}), even though the UV counterterm is in a different renormalisation scheme.

\section{IR subgraph search}
\label{sec:irsubgraphsearch}
One question that remains is how to find all IR subgraphs. Since the IR graphs could be disconnected,
it is not as straightforward as for the UV. Below we describe a method to find the complete IR spinney at once.

In section \ref{divs} the contracted IR subgraph $\tilde \gamma$ was defined by contracting the remaining 
graph (or quotient) graph $\bar \gamma=G\setminus \gamma'$ to a point in $G$, i.e.,
\beq
\tilde \gamma= G/\bar \gamma\, .
\eeq
In fact this observation generalizes further to the case of IR spinneys $S'$:
\beq
\tilde S= G/\bar S,\qquad \bar S=G\setminus S', \qquad \tilde S=\prod_i \tilde\gamma_i\,.
\eeq
The different $\tilde\gamma_i$ are then only connected through cut-vertices in $\tilde S$.
This dual description of contracted IR spinneys offers the possiblity for an alternative IR search procedure 
by searching instead for valid remaining graphs.
An easy identification of valid remaining graphs can be obtained from the contracted massless vacuum graph $G_c$ of the graph 
$G$ itself, which is defined by contracting in $G$ all the external lines in a single vertex and contracting all massive lines into points. 

All valid remaining graphs can then be identified with all spinneys of $G_c$, which include the formerly external lines. 
More precisely, we have the relation:
\beq
W'(G)=\{\tilde S\}=\{\,G_c/S | S\in W(G_c)\,, \, l_E(G) \subset S\,,\tilde\omega(G_c/S)\geq 0 \,\}\,,
\eeq
where $l_E(G)$ is the set of external lines of $G$. This allows one to construct a simple algorithm to find all IR spinneys 
by finding and combining 1PI subgraphs, similar to the construction of the UV spinney. A further advantage of this method is that disconnected IR subgraphs, such as the example 
we gave in eq.(\ref{eq:irdisconnected}), are automatically included in this alternativ search method. 

It is instructive to see how this works in an example. Consider the following graph and its associated contracted vacuum graph:
\beq
G=\IRsearch \qquad \Rightarrow \qquad G_c=\IRsearchc\;\;.
\eeq
Here we have indicated the contracted external lines in $G_c$ in green. An example for a UV spinney in $G_c$ and its associated IR spinney 
(in this case consisting of a single IR subgraph) is given by
\beq
S=\IRsearchS\qquad \Rightarrow\qquad \tilde S=G_c/S=\vactwonum{}{}{}{} \,.
\eeq
Here we used dashed lines to indicate those lines not contained in the spinney $S$. These dashed lines become the IR spinney after shrinking the disconnected 
components of $S$ to points in $G_c$.

\bibliographystyle{JHEP}
\bibliography{refs}
 
\end{document}